\documentclass[twocolumn,times]{aastex63}
\shorttitle{the mass-metallicity relation in the muse ultra deep field}
\shortauthors{Revalski et al.}
\graphicspath{{./}{figures/}}

\usepackage{float}
\usepackage{color}
\usepackage{comment}
\usepackage{graphics}
\usepackage{graphicx}
\usepackage{epstopdf}
\usepackage{microtype}
\usepackage{subfigure}
\usepackage[normalem]{ulem}
\usepackage{natbib}
\usepackage{amsmath}
\definecolor{malachite}{rgb}{0.04, 0.85, 0.32}

\usepackage{soul,xcolor}
\setstcolor{red}

\usepackage[hang, flushmargin]{footmisc}

\begin{document}

\title{\vspace{-1em}The MUSE Ultra Deep Field (MUDF). V.\\Characterizing the Mass-Metallicity Relation for Low Mass Galaxies at $\boldsymbol{z}$~$\boldsymbol{\sim}$~1\,--\,2}


\correspondingauthor{Mitchell Revalski}
\email{mrevalski@stsci.edu}

\author[0000-0002-4917-7873]{Mitchell Revalski}
\affiliation{Space Telescope Science Institute, 3700 San Martin Drive, Baltimore, MD 21218, USA}

\author[0000-0002-9946-4731]{Marc Rafelski}
\affiliation{Space Telescope Science Institute, 3700 San Martin Drive, Baltimore, MD 21218, USA}
\affiliation{Department of Physics and Astronomy, Johns Hopkins University, Baltimore, MD 21218, USA}

\author[0000-0002-6586-4446]{Alaina Henry}
\affiliation{Space Telescope Science Institute, 3700 San Martin Drive, Baltimore, MD 21218, USA}
\affiliation{Department of Physics and Astronomy, Johns Hopkins University, Baltimore, MD 21218, USA}

\author[0000-0002-9043-8764]{Matteo Fossati}
\affiliation{Dipartimento di Fisica G. Occhialini, Universit\`a degli Studi di Milano-Bicocca, Piazza della Scienza 3, I-20126 Milano, Italy}
\affiliation{INAF - Osservatorio Astronomico di Brera, via Bianchi 46, I-23087 Merate (LC), Italy}

\author[0000-0001-6676-3842]{Michele Fumagalli}
\affiliation{Dipartimento di Fisica G. Occhialini, Universit\`a degli Studi di Milano-Bicocca, Piazza della Scienza 3, I-20126 Milano, Italy}
\affiliation{INAF - Osservatorio Astronomico di Trieste, via G. B. Tiepolo 11, I-34143 Trieste, Italy}

\author[0000-0002-6095-7627]{Rajeshwari Dutta}
\affiliation{IUCAA, Postbag 4, Ganeshkind, Pune 411007, India}

\author[0000-0003-3382-5941]{Norbert Pirzkal}
\affiliation{Space Telescope Science Institute, 3700 San Martin Drive, Baltimore, MD 21218, USA}

\author[0000-0001-7396-3578]{Alexander Beckett}
\affiliation{Space Telescope Science Institute, 3700 San Martin Drive, Baltimore, MD 21218, USA}

\author[0000-0002-4770-6137]{Fabrizio Arrigoni Battaia}
\affiliation{Max-Planck-Institut f\"ur Astrophysik, Karl-Schwarzschild-Str. 1, D-85748 Garching bei M\"unchen, Germany}

\author[0000-0001-8460-1564]{Pratika Dayal}
\affiliation{Kapteyn Astronomical Institute, University of Groningen, P.O. Box 800, 9700 AV Groningen, The Netherlands}

\author[0000-0003-3693-3091]{Valentina D'Odorico}
\affiliation{INAF - Osservatorio Astronomico di Trieste, via G. B. Tiepolo 11, I-34143 Trieste, Italy}
\affiliation{Scuola Normale Superiore, Piazza dei Cavalieri 7, I-56126, Pisa, Italy}

\author[0000-0003-0083-1157]{Elisabeta Lusso}
\affiliation{Dipartimento di Fisica e Astronomia, Universit\`a di Firenze, via G. Sansone 1, I-50019 Sesto Fiorentino, Firenze, Italy}
\affiliation{INAF -- Osservatorio Astrofisico di Arcetri, Largo Enrico Fermi 5, I-50125 Firenze, Italy}

\author[0000-0001-5294-8002]{Kalina V. Nedkova} 
\affiliation{Department of Physics and Astronomy, Johns Hopkins University, Baltimore, MD 21218, USA}
\affiliation{Space Telescope Science Institute, 3700 San Martin Drive, Baltimore, MD 21218, USA}

\author[0000-0002-0604-654X]{Laura J. Prichard}
\affiliation{Space Telescope Science Institute, 3700 San Martin Drive, Baltimore, MD 21218, USA}

\author[0000-0001-7503-8482]{Casey Papovich}
\affiliation{Department of Physics and Astronomy, Texas A\&M University, College Station, TX, 77843-4242, USA}
\affiliation{George P. and Cynthia Woods Mitchell Institute for Fundamental Physics and Astronomy, Texas A\&M University, College Station, TX, 77843-4242, USA}

\author[0000-0002-4288-599X]{Celine Peroux}
\affiliation{European Southern Observatory, Karl-Schwarzschildstrasse 2, D-85748 Garching bei M{\"u}nchen, Germany}
\affiliation{Aix Marseille Universit\'e, CNRS, LAM (Laboratoire d'Astrophysique de Marseille) UMR 7326, 13388, Marseille, France}

\begin{abstract}
Using more than 100 galaxies in the MUSE~Ultra~Deep~Field with spectroscopy from the Hubble Space Telescope's Wide~Field~Camera~3 and the Very~Large~Telescope's Multi~Unit~Spectroscopic~Explorer, we extend the gas-phase mass-metallicity relation (MZR)~at~$z\approx\,$1$\,$--$\,$2 down to stellar masses of M$_{\star}$~$\approx$~10$^{7.5}$~M$_{\odot}$. The sample reaches six times lower in stellar mass and star formation rate (SFR) than previous HST studies at these redshifts, and we find that galaxy metallicities decrease to log(O/H)~+~12~$\approx$~7.8~$\pm$~0.1~(15\%~solar) at log(M$_{\star}$/M$_{\odot}$)~$\approx$~7.5, without evidence of a turnover in the shape of the MZR at low masses. We validate our strong-line metallicities using the direct method for sources with [O~III]~$\lambda$4363 and [O~III]~$\lambda$1666 detections, and find excellent agreement between the techniques. The [O~III]~$\lambda$1666-based metallicities double existing measurements with S/N~$\geq$~5 for unlensed sources at $z~>$~1, validating the strong-line calibrations up to $z~\sim$2.5. We confirm that the MZR resides $\sim$0.3~dex lower in metallicity than local galaxies and is consistent with the fundamental metallicity relation (FMR) if the low mass slope varies with SFR. At lower redshifts ($z\sim$0.5) our sample reaches $\sim$0.5~dex lower in SFR than current calibrations and we find enhanced metallicities that are consistent with extrapolating the MZR to lower SFRs. Finally, we detect only a $\sim$0.1 dex difference in the metallicities of galaxies in groups versus isolated environments. These results are based on robust calibrations and reach the lowest masses and SFRs that are accessible with HST, providing a critical foundation for studies with the Webb and Roman Space Telescopes.
\end{abstract}


\section{Introduction}\label{sec:intro}

A critical goal of extragalactic astronomy is to understand the cosmic baryon cycle, which describes how heavy elements flow through galaxies over time \citep{Peroux2020}. This includes the synthesis of heavy elements in massive stars, outflows of gas driven into the circumgalactic medium (CGM) by stellar winds, and the recycling of enriched gas into new stars via inflows from the intergalactic medium \citep{Tumlinson2017}. These complex processes produce an observational signature in the form of the mass-metallicity relation (MZR), which is a strong correlation between the stellar masses of galaxies and their gas-phase and stellar metallicities. We hereafter refer to \textit{metallicity} as the gas-phase abundance of oxygen and adopt a solar value of 12 + log(O/H)$_{\odot}$ = 8.69 $\pm$ 0.05 \citep{Asplund2009}. Characterizing the evolution of this relationship as a function of stellar mass and redshift can tightly constrain galaxy evolution models, as the shape and normalization of the MZR encapsulates how heavy elements have been produced and retained over time despite stellar feedback, gas inflows, and gas outflows \citep{Bassini2024}.

The MZR has been well-characterized in the local Universe, and was originally investigated as a function of galaxy luminosity due to difficulties in deriving accurate stellar masses \citep{Lequeux1979}. This was overcome with advances in spectral energy distribution (SED) modeling and extended to statistically significant samples using the Sloan Digital Sky Survey \citep{Tremonti2004}. Studies have now precisely characterized the MZR over the stellar mass range log(M$_{\star}$/M$_{\odot}$)~$\approx$~6~--~12 at $z\approx$~0.1 and find that metallicity is nearly constant with stellar mass at log(M$_{\star}$/M$_{\odot}$)~$>$~10, followed by a turnover and a power law decrease in metallicity at lower stellar masses \citep{Andrews2013, Berg2012, Zahid2014, Somerville2015, Ly2016, Curti2020, McQuinn2020, Cullen2021, Sanders2021, Pharo2023, He2024}.

The dispersion in the local MZR is $\sim$0.1~dex, which is remarkably small given the complex processes that govern the production, inflow, and outflow of gas within galaxies. The dispersion is further reduced when star formation rate (SFR) is included as a parameter. Specifically, galaxies with high SFRs have lower metallicities as compared to galaxies of the same stellar mass with low SFRs. At low masses, outflows driven by high SFRs may decrease the metallicity, while at the high-mass end the metallicity saturates due to a balance between self-enrichment and dilution from inflows \citep{Dayal2013}. Other factors include the dependence of SFR on the H~I gas density \citep{Lagos2016, Bothwell2013}, the rate at which stellar-driven feedback expels a significant portion of the metals, and whether inflows of pristine gas have diluted the gas reservoirs and triggered star-formation \citep{Greener2022, Yang2022, Langan2023}.

\cite{Mannucci2010} demonstrated that galaxies reside on a three-dimensional plane defined by stellar mass, metallicity, and SFR, known as the fundamental metallicity relation (FMR; \citealp{Ellison2008, Lara-Lopez2010, Lilly2013}). While still actively investigated, studies suggest that this relationship does not evolve with redshift up to at least $z\approx$~3, which suggests that similar physical mechanisms may regulate the flow of gas through galaxies over the majority of cosmic time \citep{Andrews2013, Dayal2013, Hunt2016, Gao2018, Sanders2018, Cresci2019, Kumari2021, Curti2023}.

Despite a lack of redshift evolution in the FMR, the MZR does exhibit a modest dependence on environment. Numerous studies find an enhancement in the gas-phase metallicity of galaxies in denser environments. Specifically, satellite galaxies surrounding more massive central galaxies in groups and clusters show a small but statistically significant increase in metallicity of $\sim$0.1~dex \citep{Mouhcine2007, Cooper2008, Ellison2009, Hughes2013, Kulas2013, Peng2014, Pilyugin2017, Wang2022}. This enhancement is in agreement with simulations, which find comparable or larger enhancements \citep{Wang2023}. This modest environmental dependence may be attributed to galaxy dynamics as satellite galaxies pass through the hot halos of their group \citep{Maiolino2019}.

At redshifts of $z\sim$~1 -- 3, the MZR has been characterized down to stellar masses of log(M$_{\star}$/M$_{\odot}$)~$\approx$~9 \citep{Maiolino2008, Henry2013, Kacprzak2016, Ly2016, Onodera2016, Suzuki2017, Wang2020, Sanders2021}. Studies beyond the local Universe were previously limited to higher mass galaxies due to the difficulty of observing optical emission lines at $z>$~1 from the ground that are required for metallicities \citep{Clarke2023}, combined with the intrinsic faintness of low-mass galaxies. However, the numerous population of low-mass galaxies may play the most important role in enriching the CGM, as metals can escape the weak gravitational potentials of their galaxies (\citealp{Dekel1986, D'Odorico2016, Carniani2023, Sharda2023}; see also \citealp{Baker2023} that suggest galaxy stellar mass mainly drives the relation).

Significant progress has been made in this area with large slitless spectroscopic surveys using the near-infrared capabilities of the Wide Field Camera 3 (WFC3) onboard the Hubble Space Telescope (HST). The G102 and G140 grism dispersing elements provide continuous wavelength coverage from 0.8 -- 1.7~$\mu$m at low spectral resolution, enabling the characterization of rest-frame optical diagnostic emission lines up to $z\approx$~3 (e.g. \citealp{Atek2010, Brammer2012, Momcheva2016, Pharo2019, Mowla2022, Papovich2022, Estrada-Carpenter2023, Revalski2023, Simons2021, Simons2023, Stephenson2023}). Recently, the MZR was expanded by \cite{Henry2021} down to the lowest masses characterized at these redshifts, reaching log(M$_{\star}$/M$_{\odot}$)~$\approx$~8.3 at $z\approx$~1 -- 2. Using a sample of 1056 star-forming galaxies, they found that the gas-phase metallicities are lower by 0.3~dex at $z\approx$~2 as compared to the local Universe, and confirmed the presence of a fundamental metallicity relation at these higher redshifts of $z\approx$~1 -- 2.

In following the advancements made by \cite{Henry2021}, we aim to characterize the MZR at lower stellar masses in both field and group environments. This will allow for a consistent comparison with the properties of star-forming galaxies at the peak of cosmic star formation to understand the build up of heavy elements over time. However, it is difficult to study low mass galaxies at higher redshifts without extremely sensitive imaging and spectroscopy. For this reason, we utilize recent observations of the MUSE Ultra Deep Field (MUDF) in order to characterize galaxies that are six times lower in stellar mass and SFR than earlier studies with HST at these redshifts.

The MUDF is a unique $\approx$~2\arcmin~$\times$~2\arcmin~legacy field because it contains two closely-spaced quasars at $z\approx$~3.22 that allow for galaxies observed in emission to be compared with their surrounding gas in the CGM via high resolution absorption line spectroscopy of the quasars. This field has been observed for 142 hours using the Very Large Telescope's Multi Unit Spectroscopic Explorer (VLT MUSE; ESO PID 1100.A$-$0528, PI: M.~Fumagalli; \citealp{Lusso2019, Fossati2019}), as well as for 90 orbits with HST to obtain near-infrared spectroscopy and imaging with the WFC3 (PID 15637, PI: M.~Rafelski \& M.~Fumagalli; \citealp{Revalski2023}). There are also auxiliary data sets available, including observations with VLT UVES (ESO PIDs 65.O$-$0299, 68.A$-$0216, 69.A$-$0204, 102.A$-$0194, PI: V.~D'Odorico), XMM-Newton \citep{Lusso2023}, ALMA (PID 2021.1.00285.S, PI: M.~Fumagalli), and HAWK-I (ESO PID 105.2073.001, PI: M.~Fossati).

In \cite{Revalski2023}, we described the acquisition and custom-calibration of the HST imaging and spectroscopy, and provided photometric and morphological catalogs for sources in the field. In this work, we describe our spectral fitting procedure, provide a public emission line catalog, and characterize the MZR down to the lowest stellar masses ever investigated for a sample of galaxies using HST at $z\approx$~1 -- 2. The content is organized as follows: In \S\ref{sec:data} we discuss the observations and our spectral fitting procedure. In \S\ref{sec:analysis} we describe our methodology for deriving gas-phase metallicities of individual galaxies, as well as those stacked in discrete mass intervals. In \S\ref{sec:results} we present our MZR results and explore the effects of SFR and environment. We discuss the implications of these results in \S\ref{sec:discussion} and present our conclusions in \S\ref{sec:conclusions}. We adopt the AB magnitude system ($m_\mathrm{AB}$, \citealp{Oke1983}) in this study, and assume a standard cosmology with $\Omega_M$~$\approx$~0.3, $\Omega_\Lambda$~$\approx$~0.7, and $h$~$\approx$~0.7 \citep{PlanckCollaboration2020}.

\section{Observations}\label{sec:data}

\subsection{Survey Overview}\label{ssec:data}

The survey design and data reduction for the MUSE observations (ESO PID 1100.A$-$0528, PI: M.~Fumagalli) are described in \cite{Fossati2019}, and consist of 142 hours of integral field unit spectroscopy in the optical (4600 $-$ 9350~\AA) with a spectral resolution of $R\approx$ 2000 $-$ 4000. These observations are publicly available, and calibrated data products can be retrieved using the following DOI:\dataset[10.18727/archive/84]{\doi{10.18727/archive/84}}. 

The observing strategy and custom-calibration for the HST imaging and grism spectroscopy are detailed extensively in \cite{Revalski2023}, and we provide a summary here for completeness. We obtained 90 orbits of WFC3/IR G141 grism spectroscopy (1.0 $-$ 1.7~$\mu$m) and F140W imaging in HST Cycle~26 (PID 15637, PI: M.~Rafelski \& M. Fumagalli), resulting in ultra-deep photometry and spectroscopy for $\sim$1,500 sources with a spectral resolution of $R\approx$ 150. We also utilize F336W photometry from our follow-up program (PID 15968, PI: M.~Fossati), together with archival WFPC2 F702W and F450W imaging (PID 6631, PI: P.~Francis) that provide five-filter photometry of the field for detailed SED modeling.

The default data files are available from the Mikulski Archive for Space Telescopes (MAST):\dataset[10.17909/q67p-ym16]{\doi{10.17909/q67p-ym16}}, together with our custom-calibrated data products as High Level Science Products (HLSPs) at:\dataset[10.17909/81fp-2g44]{\doi{10.17909/81fp-2g44}} and \textbf{\url{https://archive.stsci.edu/hlsp/mudf}}. In \cite{Revalski2023} we provided the photometric catalog for the field, and in this work we describe our spectral fitting procedure and publicly release the spectral emission line catalog. In the context of this study, the MUSE and HST spectra provide matched optical and near-infrared spectroscopy of $\sim$1,500 galaxies from $z \approx$~0~--~6 that enable a wide variety of science investigations. 

\subsection{Spectral Fitting}\label{ssec:fitting}

The process of simultaneously fitting ground-based and space-based spectroscopy from 4600~\AA~to 1.70~$\mu$m for thousands of sources presents several distinct challenges. We started with the emission line identification and \href{https://github.com/HSTWISP/wisp_analysis/tree/ahenry_mzr}{fitting routine} described in \cite{Henry2021} and modified it for these data to measure the redshifts and emission line fluxes of detected galaxies. First, we perform a continuous wavelet transform (CWT) on each spectrum using the \href{https://docs.scipy.org/doc/scipy/reference/generated/scipy.signal.cwt.html#scipy.signal.cwt}{CWT} feature within the \href{https://www.scipy.org}{SciPy} (v1.2.1) \href{https://docs.scipy.org/doc/scipy/reference/signal.html}{signal} package to identify sources with emission lines. We required a minimum integrated line detection of 4$\sigma$, corresponding to 2.89$\sigma$ over three continuous wavelength elements. We found that this process robustly identifies emission line sources while rejecting spectra with spurious noise and recovers 97\% of sources that were identified by visual inspection or through other automated line finding routines.

\begin{figure*}[htb]
\vspace{1em}
\centering
\includegraphics[width=\textwidth]{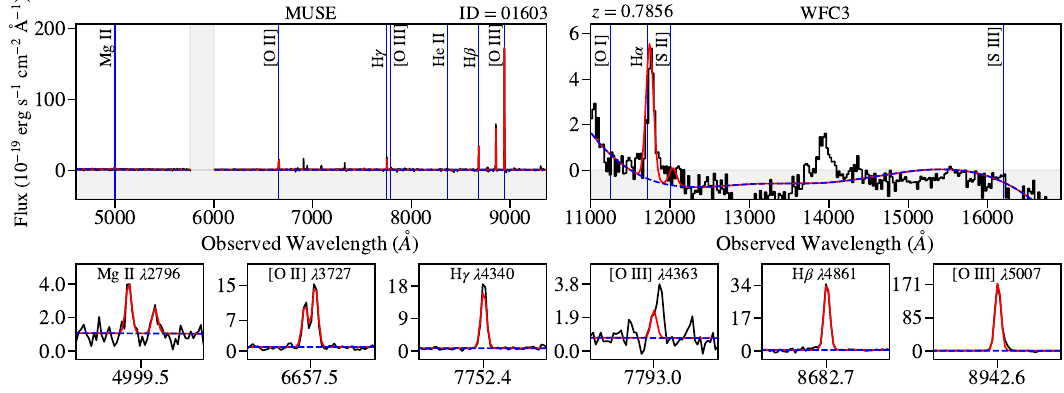}\\\vspace{2em}
\includegraphics[width=\textwidth]{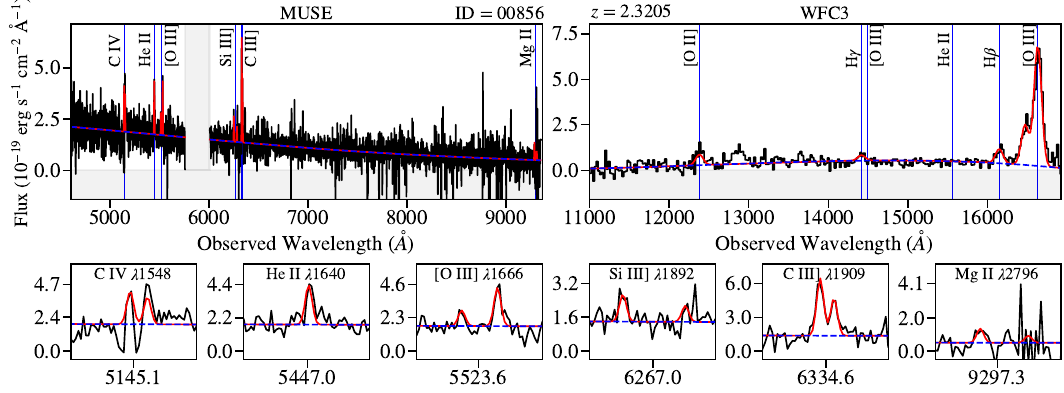}
\caption{The results produced by the fitting routine for two sources, with the MUSE spectra on the left and the WFC3 grism on the right. The data are shown in black, with Gaussian fits in red, and the derived centroids with blue vertical lines. The inset panels zoom-in on emission lines in MUSE in order of wavelength. The element and ionization state and rest wavelength for each feature are shown in the top center of each inset. There is a spectral coverage gap between MUSE and WFC3 from 9353~--~10020~\AA, while the vertical gray mask is due to the MUSE AO notch filter. These are examples of the highest S/N spectra available for the sample, which are suitable for individual metallicity measurements. Several features are notable, including the blueshifted P Cygni absorption profile of C~IV in the lower source and a weak residual skyline near [O~III] $\lambda$4363 in the upper source. The line fitting constraints allow for the effects of spurious artifacts to be minimized and flagged in the resulting fits.}
\label{fig:fitting}
\end{figure*}
\addtocounter{figure}{-1}
\begin{figure*}[htb!]
\vspace{1em}
\centering
\includegraphics[width=0.999\textwidth]{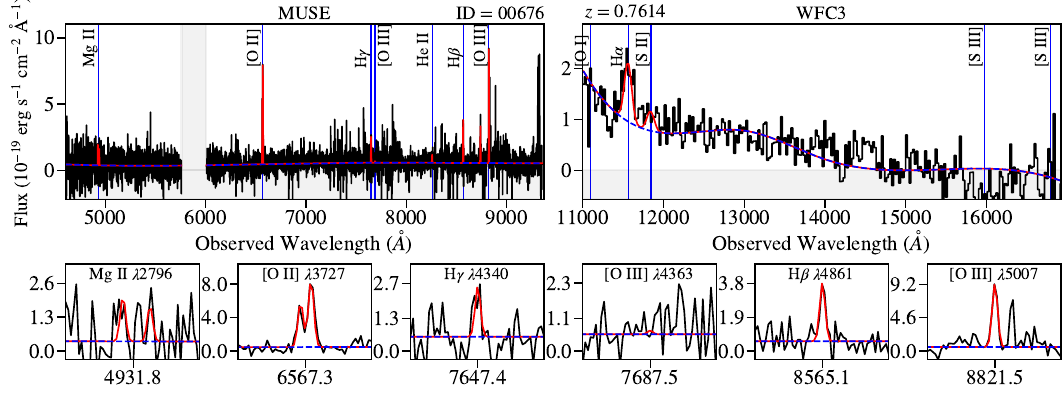}\\\vspace{2em}
\includegraphics[width=0.999\textwidth]{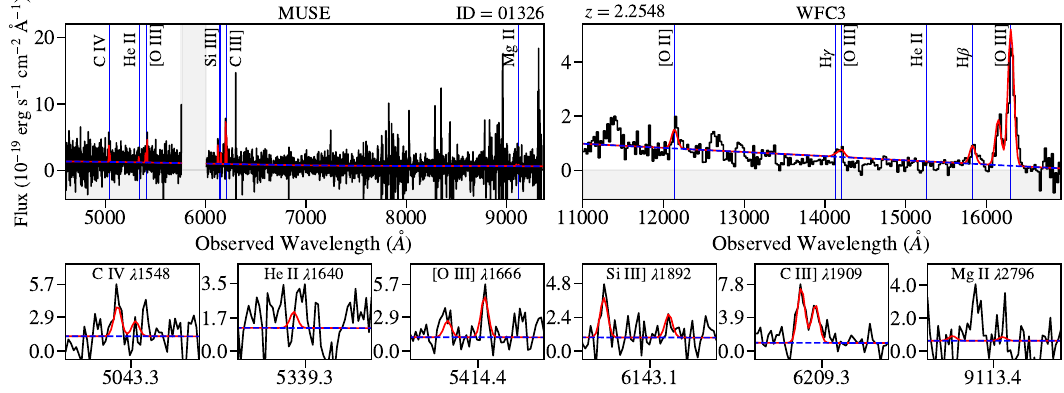}
\caption{{\it continued.} The same as above for two sources with modest S/N spectra that are representative of the typical data quality for the sample.}
\end{figure*}

We examined high signal-to-noise (S/N) spectra of the sample with strong emission lines at both low and high redshifts to identify the full range of emission features in the data, with examples shown in Figure~\ref{fig:fitting}. We identified 32 emission lines with sufficient strength to be measured in the spectra of multiple galaxies, and they are listed in Table~\ref{tab:lines}. We adopt vacuum wavelengths in units of Angstroms (\AA) throughout the analysis, but refer to lines in the text using labels common in the literature that are based on air wavelengths (e.g. [O~III] $\lambda$5007 at 5008.240~\AA). This table includes the element and ionization state, vacuum rest wavelength, energy level transition, creation ionization potential, and critical density for each emission line. While several additional weak lines are detected in a few sources, each emission line adds significant complexity and computational time in the fitting process and so are omitted from the catalog. These include H$\delta$ $\lambda$4102, [Ne~III] $\lambda\lambda$3869, 3968, H$\epsilon$ $\lambda$3971, and weaker Balmer lines. The examples in Figure~\ref{fig:fitting} highlight several key aspects of the observations. First, there is a spectral coverage gap between MUSE and WFC3 from 9353~--~10020~\AA. There is an additional small gap from 5758~--~6008~\AA~in MUSE due to the adaptive optics (AO) NaD laser notch blocking filter. These regions are masked in the spectral fitting, and are shown as gray regions in Figure~\ref{fig:fitting}. In addition, the spectral dispersion of MUSE is $\sim$17 times larger than WFC3, making the lines appear more narrow and peaked as compared to those in the WFC3 spectra. The low amplitude spikes in the MUSE spectra that do not correspond to emission lines are generally weak residual skylines that are difficult to model in the near-infrared. Overall, there are matched spectra for $\sim$1,500 galaxies.

\begin{deluxetable}{rclcc}[ht]
\setlength{\tabcolsep}{0.05in}
\def\arraystretch{1.01}
\tablecaption{Galaxy Emission Lines}
\tablehead{
\colhead{Emission} & \colhead{Vacuum} & \colhead{Transition} & \colhead{Ionization} & \colhead{Critical}\\[-0.5em]
\colhead{Line} & \colhead{Wavelength} & \colhead{Terms} & \colhead{Potential} & \colhead{Density}\\[-0.5em]
\colhead{(element)} & \colhead{(\AA)} & \colhead{(upper -- lower)} & \colhead{(eV)} & \colhead{(cm$^{-3}$)}
}
\startdata
Ly$\alpha$ & 1215.670 & 2 -- 1 & 13.6 & -- \\ \relax
N~V & 1238.821 & $^2P^\mathrm{o}_{3/2}$ -- $^2S_{1/2}$ & 77.5 & -- \\ \relax
N~V & 1242.804 & $^2P^\mathrm{o}_{1/2}$ -- $^2S_{1/2}$ & 77.5 & -- \\ \relax
C~IV & 1548.203 & $^2P^\mathrm{o}_{3/2}$ -- $^2S_{1/2}$ & 47.9 & -- \\ \relax
C~IV & 1550.777 & $^2P^\mathrm{o}_{1/2}$ -- $^2S_{1/2}$ & 47.9 & -- \\ \relax
He~II & 1640.420 & 3 -- 2 & 54.5 & -- \\ \relax
[O~III] & 1660.809 & $^5S^\mathrm{o}_{2}$ -- $^3P_1$ & 35.1 & 4.6$\times$10$^{10}$ \\ \relax
[O~III] & 1666.150 & $^5S^\mathrm{o}_{2}$ -- $^3P_2$ & 35.1 & 4.6$\times$10$^{10}$ \\ \relax
[Si~III] & 1882.707 & $^3P^\mathrm{o}_{2}$ -- $^1S_0$ & 16.4 & 4.1$\times$10$^4$ \\ \relax
Si~III] & 1892.030 & $^3P^\mathrm{o}_{1}$ -- $^1S_0$ & 16.4 & 4.2$\times$10$^{10}$ \\ \relax
[C~III] & 1906.680 & $^3P^\mathrm{o}_{2}$ -- $^1S_0$ & 24.4 & 7.7$\times$10$^4$ \\ \relax
C~III] & 1908.730 & $^3P^\mathrm{o}_{1}$ -- $^1S_0$ & 24.4 & 3.2$\times$10$^9$ \\ \relax
Mg~II & 2796.352 & $^2P^\mathrm{o}_{3/2}$ -- $^2S_{1/2}$ & 7.7 & -- \\ \relax
Mg~II & 2803.531 & $^2P^\mathrm{o}_{1/2}$ -- $^2S_{1/2}$ & 7.7 & -- \\ \relax
[O~II] & 3727.092 & $^2D^\mathrm{o}_{3/2}$ -- $^4S^\mathrm{o}_{3/2}$ & 13.6 & 4.5$\times$10$^3$ \\ \relax
[O~II] & 3729.875 & $^2D^\mathrm{o}_{5/2}$ -- $^4S^\mathrm{o}_{3/2}$ & 13.6 & 3.3$\times$10$^3$ \\ \relax
H$\gamma$ & 4341.684 & 5 -- 2 & 13.6 & -- \\ \relax
[O~III] & 4364.436 & $^1S_{0}$ -- $^1D_{2}$ & 35.1 & 2.8$\times$10$^7$ \\ \relax
He~II & 4687.020 & 4 -- 3 & 54.4 & -- \\ \relax
H$\beta$ & 4862.683 & 4 -- 2 & 13.6 & -- \\ \relax
[O~III] & 4960.295 & $^1D_{2}$ -- $^3P_1$ & 35.1 & 6.4$\times$10$^5$ \\ \relax
[O~III] & 5008.240 & $^1D_{2}$ -- $^3P_2$ & 35.1 & 6.4$\times$10$^5$ \\ \relax
[O~I] & 6302.046 & $^1D_2$ -- $^3P_2$ & 0.0 & 1.6$\times$10$^6$ \\ \relax
[O~I] & 6365.536 & $^1D_2$ -- $^3P_1$ & 0.0 & 1.6$\times$10$^6$ \\ \relax
[N~II] & 6549.850 & $^1D_2$ -- $^3P_1$ & 14.5 & 7.7$\times$10$^4$ \\ \relax
H$\alpha$ & 6564.610 & 3 -- 2 & 13.6 & -- \\ \relax
[N~II] & 6585.280 & $^1D_2$ -- $^3P_2$ & 14.5 & 7.7$\times$10$^4$ \\ \relax
[S~II] & 6718.290 & $^2D^\mathrm{o}_{5/2}$ -- $^4S^\mathrm{o}_{3/2}$ & 10.4 & 1.6$\times$10$^3$ \\ \relax
[S~II] & 6732.670 & $^2D^\mathrm{o}_{3/2}$ -- $^4S^\mathrm{o}_{3/2}$ & 10.4 & 1.5$\times$10$^4$ \\ \relax
[S~III] & 9071.100 & $^1D_2$ -- $^3P_1$ & 23.3 & 7.3$\times$10$^5$ \\ \relax
[S~III] & 9533.200 & $^1D_2$ -- $^3P_2$ & 23.3 & 7.3$\times$10$^5$ \\ \relax
He~I & 10832.86 & 4 -- 2 & 0.0 & -- \relax
\enddata
\tablecomments{Emission lines included in the spectral fitting procedure. The vacuum wavelengths are sourced from \href{https://linelist.pa.uky.edu/newpage/}{The Atomic Line List} \citep{vanHoof2018} and \cite{Leitherer2011}. The transition terms and ionization potentials required to create the emission lines are collated from \cite{Draine2011}, with the later quoted from the \href{https://www.nist.gov/pml/atomic-spectra-database}{NIST Atomic Spectra Database} \citep{NIST_ASD}. The critical densities of the forbidden emission lines are from \citet[Tables~18.1~--~18.2]{Draine2011}, while those for the semi-forbidden lines are from \cite{Wei1988}, except for Si~III] sourced from PyNeb (v1.1.17; \citealp{Luridiana2015}). The quoted critical densities depend strongly on the adopted atomic data, with some values in the literature differing by up to factors of three.}
\label{tab:lines}
\end{deluxetable}

\begin{deluxetable}{rclc}[ht]
\setlength{\tabcolsep}{0.06in}
\tabletypesize{\small}
\tablecaption{Spectral Fitting Constraints}
\tablehead{
\colhead{Emission Line} & \colhead{} & \colhead{Relative Ratio} & \colhead{Initial Value}
}
\startdata
\textbf{Constant} &  &  & \\ \hline
[O~III] $\lambda$1666 & = & 2.46 $\times$ [O~III] $\lambda$1660 & N/A\\ \relax
[O~III] $\lambda$5007 & = & 3.00 $\times$ [O~III] $\lambda$4959 & N/A\\ \relax
[O~I] $\lambda$6300 & = & 3.00 $\times$ [O~I] $\lambda$6363 & N/A\\ \relax
[N~II] $\lambda$6585 & = & 3.00 $\times$ [N~II] $\lambda$6550 & N/A\\ \relax
[S~III] $\lambda$9533 & = & 2.48 $\times$ [S~III] $\lambda$9071 & N/A\\ \hline \relax 
\textbf{Variable} &  &  & \\ \hline
N~V $\lambda$1239 & = & 1.15 -- 2.00 $\times$ N~V $\lambda$1243 & 1.95\\ \relax
C~IV $\lambda$1548 & = & 1.22 -- 2.00 $\times$ C~IV $\lambda$1551 & 1.98\\ \relax
[Si~III] $\lambda$1883 & = & 5E-5 -- 1.69 $\times$ Si~III] $\lambda$1892 & 0.50\\ \relax
[C~III] $\lambda$1907 & = & 1E-5 -- 1.54 $\times$ C~III] $\lambda$1909 & 0.92\\ \relax
Mg~II $\lambda$2796 & = & 1.26 -- 2.00  $\times$ Mg~II $\lambda$2804 & 1.98\\ \relax
[O~II] $\lambda$3727 & = & 0.66 -- 2.63 $\times$ [O~II] $\lambda$3730 & 0.66 \\ \relax
[S~II] $\lambda$6716 & = & 0.42 -- 1.47 $\times$ [S~II] $\lambda$6731 & 1.40 \relax
\enddata
\tablecomments{The relative flux ratios of emission line doublets that were constrained to exact values (upper rows) or specific ranges (lower rows) in the spectral fits. The fixed ratios are based on atomic transition probabilities, while those with ranges are based on Cloudy photoionization models (see text). In the case of variable ratios, the initial value used in the fitting process is listed and typically represents the low density limit. The [Si III], [C III], [O~II], and [S~II] doublets are all sensitive to the electron density of the gas.}
\label{tab:lines2}
\vspace{-2em}
\end{deluxetable}

Next, we fit the spectra for each object with detected emission lines. As briefly described in \cite{Revalski2023}, a cubic spline is first fit to the continuum of the WFC3 and MUSE spectra simultaneously, making an initial guess of the source redshift based on the strongest emission line. We can then easily change the guess to other emission lines, adjust the wavelength regions used in the continuum fit, mask contaminated regions, and then choose to reject the fit or save the results for each spectral fit to an emission line catalog. We place constraints on the model parameters to converge on a successful fit and implement physical limits. First, the widths of emission lines in WFC3 with a dispersion of 21.5~\AA~pixel$^{-1}$, relative to those in MUSE with a dispersion of 1.25~\AA~pixel$^{-1}$, are initially fixed at a ratio of 17.2. However, sources that are physically extended in the grism observations have larger line spread functions, and we found that allowing this ratio to increase by up to a factor of two for the most extended galaxies produced excellent fits. We thus constrained the relative widths by this factor when fitting the full sample of galaxies.

Next, we implement limits on the relative ratios of emission line doublets based on atomic physics and Cloudy photoionization models \citep{Ferland2017}. Specifically, doublets with the same upper energy level and closely spaced lower levels have relative intensities that are constant. In addition, doublets with closely spaced upper levels and the same lower energy state that have different transition probabilities (or critical densities) are sensitive to the electron density of the gas. In these cases, we use a grid of Cloudy models over a broad range of 9~dex in density and ionization from \cite{Revalski2018} to constrain the doublet ratios between their low and high density limits. These grids were generated for a hydrogen column density of N$_\mathrm{H}$ = 10$^\mathrm{21.5}$~cm$^{-2}$ without dust. A lack of dust in the models is suitable for low mass galaxies \citep{Shapley2022, Shapley2023}, but generally has a negligible effect on the relative ratios of the same ionic species. The ratios of emission lines with fixed relative intensities, as well as those constrained to specific ranges, are provided in Table~\ref{tab:lines2}.

In addition, all of the emission lines must yield a consistent redshift, but modest deviations are allowed for specific pairs of lines due to small differences in wavelength calibration between the two spectrographs, as well as radiative transfer effects that can result in offsets between emission lines of different ions. We impose the following constraints on eight groups of lines, requiring that each group of lines have the same redshift: 1) Ly$\alpha$ $\lambda$1216 and N~V $\lambda\lambda$1239, 1243; 2) He~II $\lambda$1640, [O~III] $\lambda\lambda$1660, 1666, [Si~III] $\lambda\lambda$1883, 1892, and [C~III] $\lambda\lambda$1907, 1909; 3) H$\gamma$ $\lambda$4342, [O~III] $\lambda$4363, He~II $\lambda$4687, H$\beta$ $\lambda$4862, and [O~III] $\lambda\lambda$4959, 5007; 4) [O~I] $\lambda\lambda$6300, 6363, [N~II] $\lambda\lambda$6550, 6585, H$\alpha$ $\lambda$6565, and [S~II] $\lambda\lambda$6716, 6731; 5) [S~III] $\lambda\lambda$9071, 9533, and He~I $\lambda$10832. The remaining three constraints are for C~IV $\lambda\lambda$1548, 1551, Mg~II $\lambda\lambda$2796, 2804, and [O~II] $\lambda\lambda$3727, 3730, which are independent of the other lines except for when Mg~II or [O~II] are in the grism spectral range at higher redshifts, in which cases they are fixed to the other emission lines seen in the grism.

These sets of lines are allowed to vary from one another by up to $\delta z$ = 0.00334 (1000 km~s$^{-1}$), which produces visually excellent fits for 98\% of the galaxies, with lines for 83\% of galaxies offset by $\leq$~500 km~s$^{-1}$ (see Figure~11 in \citealp{Revalski2023}). In the remaining few objects, using twice this limit produced visually acceptable fits. Given the different spectral resolutions, calibration systematics between ground and space-based observations, and radiative transfer effects on lines of different ionic species, this agreement is excellent.

Furthermore, emission lines close in wavelength are often blended in the low dispersion WFC3 grism observations, requiring additional constraints for successful fitting. Specifically, the H$\alpha$ $\lambda$6565 and [N~II] $\lambda\lambda$6550, 6585 emission lines are fully blended and cannot be deconvolved \citep{Henry2021}. In general, the [N~II] emission lines are weak in star-forming galaxies and following the model of \cite{Henry2021}, we fix the flux of [N~II] $\lambda$6585 to 10\% of H$\alpha$ in the fitting procedure when these lines are in the grism ($z >$~0.52). We note that this constraint does not apply to the metallicity analysis discussed later, where the contribution from [N~II] is self-consistently forward modeled to determine metallicities in a Bayesian framework. This constraint corresponds to 13\% of the total flux including both [N~II] lines. This overall contribution is in good agreement with the values presented in Table~2 of \cite{Erb2006} for moderate mass galaxies, where the [N~II] contribution ranges from 6 -- 22\%, increasing with galaxy stellar mass. When these lines are observed in the MUSE spectra, their fluxes are free to vary independently within the constraints listed in Table~\ref{tab:lines2}. This degree of freedom allows us to confirm the validity of the constraint at higher redshifts, and we used 30 lower mass galaxies in MUSE with [N~II] detections at S/N $>$ 5 to confirm that, on average, the [N~II] doublet contributes 12\% of the combined flux with H$\alpha$.

Finally, the Ly$\alpha$ $\lambda$1216 emission line is often asymmetric, displaying a strong wing at longer wavelengths due to radiative transfer effects \citep{Verhamme2006, Dijkstra2007, Laursen2010, Childs2018, Hu2023}. We adopt a simple approach \citep{Hu2010} for the continuum-detected Ly$\alpha$ emitters and fit a truncated half-Gaussian with the same centroid as the main Gaussian component to account for this wing emission. While directly integrating under the line profile may produce a more precise estimate of the total Ly$\alpha$ flux, this parametric approach allows the full spectrum to be fit self-consistently. In addition, Ly$\alpha$ nebulae are known to have larger extents than the general nebular emission lines as emission diffuses outwards to larger radii such that an aperture of fixed size will miss a substantial portion of the emission \citep{Leclercq2017, Fossati2019, Urrutia2019}. Considering these factors, we fit the Ly$\alpha$ line in our spectra so it will not affect the continuum model and adjacent emission lines, but will provide the measured fluxes in a subsequent publication using dedicated apertures to properly extract the extended Ly$\alpha$ emission.

In many cases, the Ly$\alpha$, Mg~II, and C~IV lines show interesting systematic offsets relative to the other emission lines, which is expected due to resonant scattering \citep{Henry2018, Berg2019}. The C~IV lines specifically are created by nebular emission, stellar winds, and are also seen in absorption in the ISM. These lines are susceptible to complex optical depth effects where photons near the systemic velocity encountering high optical depth are scattered into the wings of the lines where the optical depth is lower. As such, we recommend specialized fitting of these UV emission lines beyond that completed in this study for users interested in the detailed physics, complex line profiles, and precise fluxes of these UV diagnostic lines. In cases where strong absorption causes deviations from the line ratio constraints listed in Table~\ref{tab:lines2} (e.g. C~IV in Figure~\ref{fig:fitting}), the line flux measurements are generally flagged as unreliable in the spectral catalog (see Table~\ref{tab:catalog}).

Using this framework, we successfully derived spectroscopic redshifts for 419 sources that were published in the publicly-available source catalog \citep{Revalski2023}\footnote{In addition to fitting emission lines, eight of these redshifts were derived from strong absorption features using \url{https://matteofox.github.io/Marz} and \url{https://github.com/mifumagalli/mypython}.}. Those redshifts are duplicated in the current emission line catalog presented here for completeness. By default the redshift measurements are based on H$\alpha$~$\lambda$6565, and when the well-resolved [O~II]~$\lambda$$\lambda$3727, 3730 doublet is available in MUSE ($z<$~1.5) we apply the \verb|o2_3727_dz| offset value in the catalog (see Table~\ref{tab:catalog}) to take advantage of MUSE's superior spectral resolution and obtain the most precise redshifts. The fitting procedure does not include a model for any underlying stellar absorption around the Balmer emission lines, which we account for later in our metallicity calculations.

In Table~\ref{tab:catalog}, we list the columns that are included in the emission line catalog. These include the object ID, redshift, sky coordinates, magnitude, size measurements, spectral fit parameters, and their associated uncertainties. These are followed by the integrated line flux and error, both in units of erg s$^{-1}$ cm$^{-2}$, the observed equivalent width (EW) in \AA~(which is negative when the continuum is not detected), and a contamination flag to indicate the quality of the fit for each emission line. These flags have values of 0 -- 5, with the flag for an excellent fit being 0. A slightly contaminated fit has a flag of 1, and a poor fit that should not be used has a flag of 5. A value of 3 indicates a non-ideal continuum fit such that the fluxes may be suspect, and flags are added in a bit-wise manner. Only measurements with a flag of 0, or a flag of 1 after visual inspection, should be used for science investigations to ensure the cleanest samples free from fitting and data artifacts. We note that we also fit the spectra of the two primary quasars (IDs 1535 and 20405) to determine their redshifts, but their line fluxes are flagged and should not be used for investigations. These quasar spectra contain complex broad and narrow emission and absorption features, as well as asymmetric line profiles and large centroid shifts that require detailed modeling that is beyond the scope of this study.

Finally, the individual and integrated fluxes of closely spaced doublet lines are both saved in the catalog. As examples, the [O~II]~$\lambda$$\lambda$3727, 3730 and [S~II]~$\lambda$$\lambda$6716, 6731 emission line doublets are resolved in MUSE, which allows their individual components to be used for density diagnostics. However, these lines are fully blended in the WFC3 data and so only the total line flux should be used. In general, the individual components and their sums can be used when present in the MUSE data, and only the summed flux should be used when the lines are present in WFC3 at higher redshift (e.g. H$\alpha$ + [N~II] and [S~II] at $z >$~0.5, and [O~II] at $z >$~1.7).


\startlongtable
\begin{deluxetable}{cll}
\vspace{-1.925em}
\def\arraystretch{1.22}
\setlength{\tabcolsep}{0.025in} 
\tablecaption{Emission Line Catalog Columns}
\tablehead{
\colhead{Column \#} & \colhead{Parameter} & \colhead{Description}
}
\startdata
(1) & objid & Object ID\\ 
(2) & redshift & Redshift value\\ 
(3) & redshift\_error & Redshift uncertainty\\ 
(4) & ra\_obj & Right Ascension\\ 
(5) & dec\_obj & Declination\\ 
(6) & f140w\_mag & $m_\mathrm{F140W}$ magnitude\\ 
(7) & a\_image\_obj & Semi-major axis size (pixels)\\ 
(8) & b\_image\_obj & Semi-minor axis size (pixels)\\ 
(9) & snr\_tot\_others & Flux-weighted SNR of lines\\ 
(10) & chisq & Chi-Square of spectral fit\\ 
(11) & fwhm\_muse & FWHM of lines in MUSE\\ 
(12) & fwhm\_muse\_error & FWHM uncertainty\\ 
(13) & fwhm\_g141 & FWHM of lines in WFC3\\ 
(14) & fwhm\_g141\_error & FWHM uncertainty\\ 
(15) & la\_1216\_dz & $\delta z$ for lines locked to Ly$\alpha$\\ 
(16) & c4\_1548\_dz & $\delta z$ for lines locked to C~IV\\ 
(17) & uv\_line\_dz & $\delta z$ for lines locked to He~II\\ 
(18) & m2\_2796\_dz & $\delta z$ for Mg~II\\ 
(19) & o2\_3727\_dz & $\delta z$ for [O~II]\\ 
(20) & o3\_5007\_dz & $\delta z$ for lines locked to [O~III]\\ 
(21) & s3\_he\_dz & $\delta z$ for lines locked to [S~III]\\ \hline
(22-25) & la\_1216\_$^{\star}$ & flux, error, ew\_obs, contam\\ \relax
(26-29) & la\_wing\_$^{\star}$ & flux, error, ew\_obs, contam\\ \relax
(30-33) & la\_1216\_wing\_$^{\star}$ & flux, error, ew\_obs, contam\\ \relax
(34-37) & n5\_1238\_$^{\star}$ & flux, error, ew\_obs, contam\\ \relax
(38-41) & n5\_1242\_$^{\star}$ & flux, error, ew\_obs, contam\\ \relax
(42-45) & n5\_1238\_1242\_$^{\star}$ & flux, error, ew\_obs, contam\\ \relax
(46-49) & c4\_1548\_$^{\star}$ & flux, error, ew\_obs, contam\\ \relax
(50-53) & c4\_1550\_$^{\star}$ & flux, error, ew\_obs, contam\\ \relax
(54-57) & c4\_1548\_1550\_$^{\star}$ & flux, error, ew\_obs, contam\\ \relax
(58-61) & h2\_1640\_$^{\star}$ & flux, error, ew\_obs, contam\\ \relax
(62-65) & o3\_1660\_$^{\star}$ & flux, error, ew\_obs, contam\\ \relax
(66-69) & o3\_1666\_$^{\star}$ & flux, error, ew\_obs, contam\\ \relax
(70-73) & o3\_1660\_1666\_$^{\star}$ & flux, error, ew\_obs, contam\\ \relax
(74-77) & s3\_1883\_$^{\star}$ & flux, error, ew\_obs, contam\\ \relax
(78-81) & s3\_1892\_$^{\star}$ & flux, error, ew\_obs, contam\\ \relax
(82-85) & s3\_1883\_1892\_$^{\star}$ & flux, error, ew\_obs, contam\\ \relax
(86-89) & c3\_1907\_$^{\star}$ & flux, error, ew\_obs, contam\\ \relax
(90-93) & c3\_1909\_$^{\star}$ & flux, error, ew\_obs, contam\\ \relax
(94-97) & c3\_1907\_1909\_$^{\star}$ & flux, error, ew\_obs, contam\\ \relax
(98-101) & m2\_2796\_$^{\star}$ & flux, error, ew\_obs, contam\\ \relax
(102-105) & m2\_2803\_$^{\star}$ & flux, error, ew\_obs, contam\\ \relax
(106-109) & m2\_2796\_2803\_$^{\star}$ & flux, error, ew\_obs, contam\\ \relax
(110-113) & o2\_3727\_$^{\star}$ & flux, error, ew\_obs, contam\\ \relax
(114-117) & o2\_3730\_$^{\star}$ & flux, error, ew\_obs, contam\\ \relax
(118-121) & o2\_3727\_3730\_$^{\star}$ & flux, error, ew\_obs, contam\\ \relax
(122-125) & hg\_4342\_$^{\star}$ & flux, error, ew\_obs, contam\\ \relax
(126-129) & o3\_4363\_$^{\star}$ & flux, error, ew\_obs, contam\\ \relax
(130-133) & h2\_4686\_$^{\star}$ & flux, error, ew\_obs, contam\\ \relax
(134-137) & hb\_4863\_$^{\star}$ & flux, error, ew\_obs, contam\\ \relax
(138-141) & o3\_4959\_$^{\star}$ & flux, error, ew\_obs, contam\\ \relax
(142-145) & o3\_5007\_$^{\star}$ & flux, error, ew\_obs, contam\\ \relax
(146-149) & o3\_4959\_5007\_$^{\star}$ & flux, error, ew\_obs, contam\\ \relax
(150-153) & o1\_6300\_$^{\star}$ & flux, error, ew\_obs, contam\\ \relax
(154-157) & o1\_6363\_$^{\star}$ & flux, error, ew\_obs, contam\\ \relax
(158-161) & o1\_6300\_6363\_$^{\star}$ & flux, error, ew\_obs, contam\\ \relax
(162-165) & n2\_6550\_$^{\star}$ & flux, error, ew\_obs, contam\\ \relax
(166-169) & ha\_6565\_$^{\star}$ & flux, error, ew\_obs, contam\\ \relax
(170-173) & n2\_6585\_$^{\star}$ & flux, error, ew\_obs, contam\\ \relax
(174-177) & ha\_6550\_6565\_6585\_$^{\star}$ & flux, error, ew\_obs, contam\\ \relax
(178-181) & s2\_6716\_$^{\star}$ & flux, error, ew\_obs, contam\\ \relax
(182-185) & s2\_6731\_$^{\star}$ & flux, error, ew\_obs, contam\\ \relax
(186-189) & s2\_6716\_6731\_$^{\star}$ & flux, error, ew\_obs, contam\\ \relax
(190-193) & s3\_9069\_$^{\star}$ & flux, error, ew\_obs, contam\\ \relax
(194-197) & s3\_9532\_$^{\star}$ & flux, error, ew\_obs, contam\\ \relax
(198-201) & s3\_9069\_9532\_$^{\star}$ & flux, error, ew\_obs, contam\\ \relax
(202-205) & he10830\_$^{\star}$ & flux, error, ew\_obs, contam \relax
\enddata
\tablecomments{A list of the columns provided in the emission line catalog. Entries include the object ID, redshift, sky coordinates, magnitude, size measurements, spectral fit parameters, and their associated uncertainties. The $\delta z$ values are offsets relative to the redshift of H$\alpha$. These are followed by entries for each emission line with the flux and flux error in units of erg s$^{-1}$ cm$^{-2}$, observed equivalent width in \AA, and a quality flag (see text). The individual and total fluxes of closely spaced doublets are saved in the catalog so the component fluxes can be used when the lines are resolved in MUSE, and the integrated fluxes can be used when lines are blended in WFC3. The fluxes of non-detected lines are recorded with a value of --1.}
\label{tab:catalog}
\end{deluxetable}

\section{Analysis}\label{sec:analysis}

\subsection{Stellar Population Synthesis Models}\label{ssec:sps}

We require a robust stellar mass and star-formation rate for each galaxy to investigate their location in the MZR and FMR, which we calculate using stellar population synthesis (SPS) models. The stellar mass and star-formation rate estimates are obtained by simultaneously fitting the multiwavelength photometry and the MUSE spectra (when available) with the Monte Carlo Spectro-Photometric Fitter (MC-SPF) code \citep{Fossati2018}. The details of the fitting procedure are given in \cite{Fossati2019} and Fossati et al. (in preparation). In summary, we start by fitting the data with \cite{Bruzual2003} models obtained with a \cite{Chabrier2003} initial mass function (IMF) and an exponentially declining star-formation history (SFH). The models include emission lines using line ratios from \cite{Byler2017} that are scaled to the Lyman continuum luminosity of the stellar templates. We also include dust attenuation with a double \cite{Calzetti2000} attenuation law where the flux from stars older than 10~Myr are attenuated with a curve that is normalised by a free parameter ($A_V$) in the fitting procedure, while the same curve is normalised by 2.27 $\times A_V$ for stars younger than 10~Myr. While SFRs can also be estimated directly from the Balmer emission lines, H$\alpha$ is blended with [N~II] in the grism, it falls outside of our wavelength coverage at higher redshifts ($z >$~1.6), and the Balmer lines require a correction for stellar absorption, so adopting the SPS model parameters allows us to self-consistently model galaxies across all redshifts.


While it is common to adopt solar metallicity for the stellar component, this assumption is likely not appropriate for all of our galaxies that span a large dynamic range in redshift and stellar mass. We use our multi-filter photometry and high S/N spectroscopy to model the metallicity of the stars in discrete metallicity bins. The \cite{Bruzual2003} models provide stellar templates at 100\% solar, 40\% solar and 20\% solar metallicity. We first calculate the stellar masses using stellar and emission line templates at solar metallicity. Next, we refine the SPS metallicity in one of two ways. First, for sources with an individual gas-phase metallicity measurement, we use the closest stellar template metallicity value. For all other sources, we compare the mass of the target with our stacked MZRs in the appropriate redshift bin (above and below $z$~=~1), and recalculate the SPS models using a lower stellar metallicity if the object is in the low mass regime. We iteratively evaluate the template metallicity based on the new value of the stellar mass until all galaxies are assigned into one of the three broad metallicity bins provided by the \cite{Bruzual2003} models (100\%, 40\%, or 20\% solar metallicity). The effect on the derived masses is smallest at low masses and increasingly important for higher mass galaxies. We only utilize stellar masses and SFRs obtained with this procedure.

\subsection{Sample Selection}\label{ssec:sample}

We minimally require spectra with detections of the [O~II] $\lambda\lambda$3727, 3730, H$\beta$ $\lambda$4862, and [O~III] $\lambda\lambda$4959, 5007 emission lines to calculate metallicities (see \S\ref{ssec:calibrations}). An ideal survey would have very high S/N spectra such that we could derive metallicities for each galaxy. However, in practice it is necessary to implement S/N criteria on the emission lines for individual sources, which can bias the sample towards galaxies with higher SFRs that produce stronger emission lines. This can bias the measurements of lines near the S/N limit, as the larger number of weakly emitting sources below the threshold have a higher chance of being scattered into the sample than strong line emitters being scattered out of the sample due to the measurement uncertainties \citep{Eddington1913}.

Alternatively, spectra that share common properties can be stacked together to create high S/N composites \citep{Henry2013, Wang2018}. This reduces noise and leads to better constraints on the average emission line properties of galaxies, but this can also bias the resulting measurements if all of the galaxies do not share similar properties \citep{Andrews2013}. Importantly, by stacking spectra we reduce the measurement scatter in the MZR and robustly determine the metallicities of galaxies down to lower masses than is possible for individual sources. By utilizing the very sensitive spectroscopy available for the MUDF, we minimize biases due to SFR selection by reaching SFRs of $\sim$1~M$_{\odot}$ yr$^{-1}$ for stellar masses of M$_{\star}$~$\lesssim$~10$^8$~M$_{\odot}$ at $z\approx$~1 -- 2. Using these observations, we investigate the average MZR for composite spectra produced by stacking galaxies with similar masses, as well as for individual galaxies with very high S/N spectroscopy.

We use our emission line catalog to select star-forming galaxies from the 419 sources with spectroscopic redshifts. First, we require that the [O~III] $\lambda\lambda$4959, 5007 emission line doublet is detected at S/N~$\geq$~3, which results in 249 sources. Next, we use the Mass-Excitation (MEx) diagram developed by \cite{Juneau2011, Juneau2014} to remove active galactic nuclei (AGN) from the sample. This diagnostic is similar to traditional Baldwin–Phillips–Terlevich (BPT) diagrams \citep{Baldwin1981, Veilleux1987, Kewley2001}, except the redder of the two emission line ratios (e.g. [N~II]/H$\alpha$) is replaced with stellar mass (M$_{\star}$), enabling it to be used at higher redshifts and lower spectral resolution.

\begin{figure}[t]
\centering
\includegraphics[width=\columnwidth]{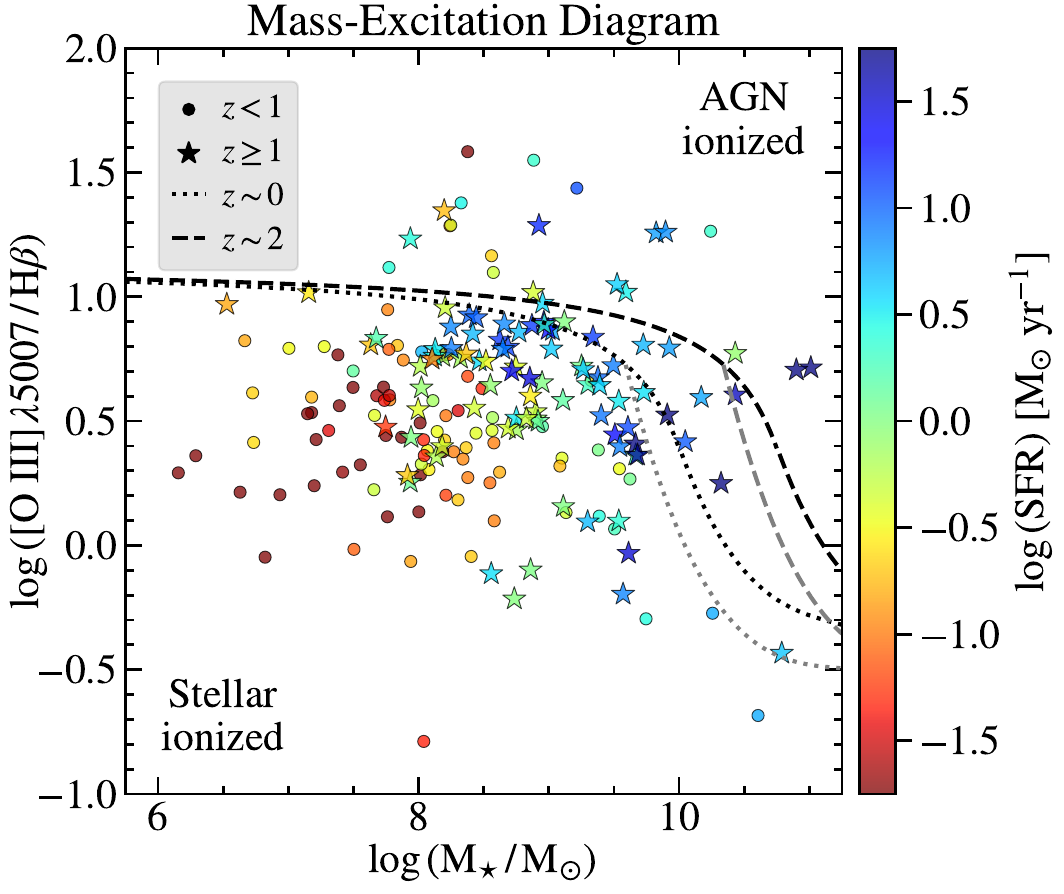}\\
\vspace{-0.5em}
\caption{The mass-excitation (MEx) diagram for 243 galaxies where the [O~III] $\lambda$5007~\AA~emission line is detected at S/N~$\geq$~3. Sources at $z<$~1 are shown with circles, while sources at $z\geq$~1 are shown with stars, with points color-coded based on their SFR from SED modeling. The demarcation lines separating stellar-ionized and AGN-ionized emission line sources at $z\approx$~0 (dotted lines) are from \cite{Juneau2014}, and are also shown shifted by 0.75 dex to higher mass for use at $z\approx$~2 (dashed lines). The area between the gray and black lines represents a composite region with contributions from AGN and star-formation. The $z<$~1 sources (circles) should be compared with the dotted lines, while the $z\geq$~1 sources (stars) are compared with the dashed lines. This diagnostic identifies 20 sources as AGN that we exclude from the metallicity analysis.}
\label{fig:mex}
\end{figure}

We show the MEx diagram for our sample in Figure~\ref{fig:mex}, where the demarcation lines separating stellar-ionized and AGN-ionized emission line sources at $z\approx$~0 from \cite{Juneau2014} are also shown shifted by 0.75 dex to higher masses for use at $z\approx$~2. As described by \cite{Coil2015}, this redshift evolution occurs because galaxies at higher redshifts have lower metallicities, resulting in higher [O~III]/H$\beta$ ratios at fixed stellar mass. Without this correction, the high mass $z \geq$~1 star-forming galaxies that reside between these two regions would be incorrectly flagged as AGN. This diagnostic uses only the $\lambda$5007 component of the [O~III] doublet, and including the $\lambda$4959 line, which is always 1/3 the strength, would require shifting the demarcation lines to higher [O~III]/H$\beta$ by log(1 + 1/3) $\approx$ 0.125~dex. This diagnostic identifies 20 sources as AGN that we exclude from the analysis because the abundance methods are calibrated for stellar-ionized gas. In addition, we take advantage of the rest-frame UV coverage at higher redshifts to select sources where the He~II~$\lambda$1640, C~III]~$\lambda$1907, 1909, and O~III] $\lambda$1666 emission lines are detected at S/N~$\geq$~4 to construct the C3He2-O3He2 ionization diagram discussed in \cite{Mingozzi2023} and \cite{Feltre2016}. There are 11 sources meeting these criteria, and all are consistent with star-forming galaxies. Finally, the XMM-Newton sources reported by \cite{Lusso2023} yield no additional AGN beyond those identified in the MEx diagram, and we omit four more sources that have insufficient photometry required to calculate accurate stellar mass models.

\begin{figure*}[htb!]
\centering
\includegraphics[width=\textwidth]{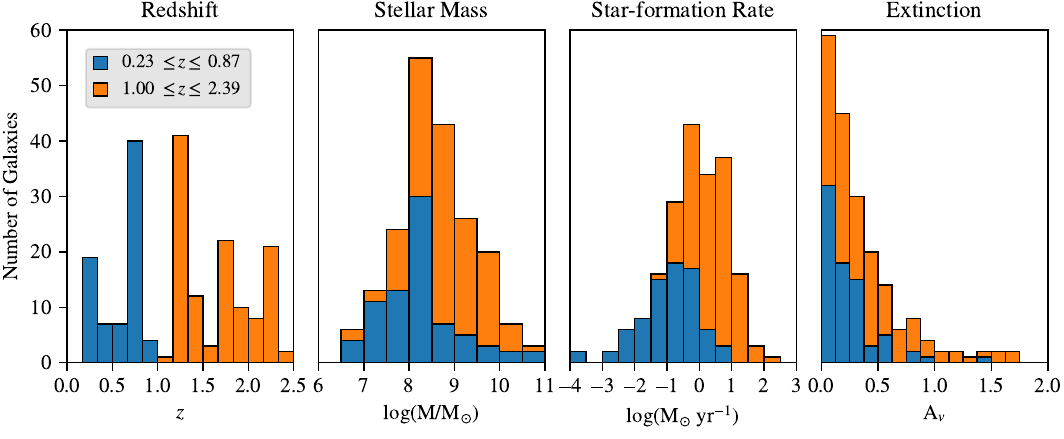}\\
\vspace{-1em}
\caption{The redshifts, stellar masses, SFRs, and extinctions for the 197 sources with coverage of the [O~II], [O~III], and H$\beta$ emission lines. These galaxies are divided into a low redshift sample of 77 galaxies shown in orange, and a high redshift sample of 120 galaxies shown in blue.}
\label{fig:sample}
\end{figure*}

These selection criteria result in 225 sources at $z\leq$~2.39, which is the highest redshift at which we can detect [O~III] $\lambda$5007 in WFC3 at its maximum wavelength limit of 1.70~$\mu$m. There are also four additional limiting redshift windows: [O~II] is blueward of the MUSE spectral range at $z<$~0.23 and falls in the adaptive optics NaD laser notch blocking filter at $z=$~0.54~--~0.61; [O~III] and/or H$\beta$ reside in the wavelength gap between MUSE and WFC3 (9353~--~10020~\AA) at $z=$~0.87~--~1.06; and [O~II] resides in the wavelength gap at $z=$~1.51~--~1.69. These redshift constraints yield 197 sources at $z=$~0.23~--~2.39 with wavelength coverage of [O~II], [O~III], and H$\beta$. Of these, 115 have non-zero fluxes of all three lines that are required for individual metallicity measurements. The properties of these 197 galaxies are summarized in Figure~\ref{fig:sample}.

\subsection{Subsample Construction}

We use the sources with wavelength coverage of [O~II], [O~III], and H$\beta$ to construct subsamples for analysis. Specifically, from the full sample described above we select 72 galaxies at $z=$~0.23 -- 0.87 and 109 galaxies at $z=$~1.00 -- 2.39 with S/N~$\geq$~5 in [O~III] $\lambda\lambda$4959, 5007. We refer to these as our low and high redshift samples, respectively. In the low redshift sample, [O~II], [O~III], and H$\beta$ are observed in the MUSE spectra, while for the high redshift sample [O~III] and H$\beta$ are redshifted into WFC3. Primarily, we are interested in characterizing the MZR at $z\approx$~1 -- 2 over the same mass range that has been studied in the local Universe. The galaxies in our high redshift sample have high quality spectra with a median S/N~$\approx$~17 for the [O~III] doublet and S/N~$\approx$~2.5 for H$\beta$. Our requirment of detecting [O~III] at S/N~$\geq$~5 ensures that only sources with robust redshifts are selected and they will contribute meaningful information to the spectral stacks. We apply the \verb|o3_5007_dz| offset from our catalog so the redshifts are based on the strong [O~III] $\lambda$5007 emission line.

We divide this subsample into six mass bins from log(M$_{\star}$/M$_{\odot}$)~$=$~6.5~--~11.0, with each bin having a width of 0.5~dex in stellar mass, except the highest and lowest mass bins that span $\sim
$1~dex to encapsulate enough galaxies for useful statistics. Similarly for the low redshift sample, we divide the galaxies into six mass bins from log(M$_{\star}$/M$_{\odot}$)~$=$~6.5~--~11.0, with each bin having a width of 0.5~dex in stellar mass, except the highest bin that encapsulates all galaxies at log(M$_{\star}$/M$_{\odot}$)~$>$~9.5. Next, we create a median stack of the spectra in each bin following the procedures of \cite{Henry2013}. First, the spectra are continuum-subtracted using the results from our spectral fits, normalized to their [O~III] fluxes, and de-redshifted to rest wavelengths. A median spectrum is then calculated for each mass bin, and the uncertainties are determined using bootstrap resampling \citep{Henry2021}. In this process, we resample the MUSE spectra to the spectral resolution of WFC3 by smoothing the MUSE spectra using a Gaussian kernel with a width equal to the WFC3 spectral resolution, and then apply a flux-conserving resampler \citep{Carnall2017} to match the spectrum to the WFC3 wavelength grid.

\begin{figure*}[htb!]
\centering
\includegraphics[width=0.88\textwidth]{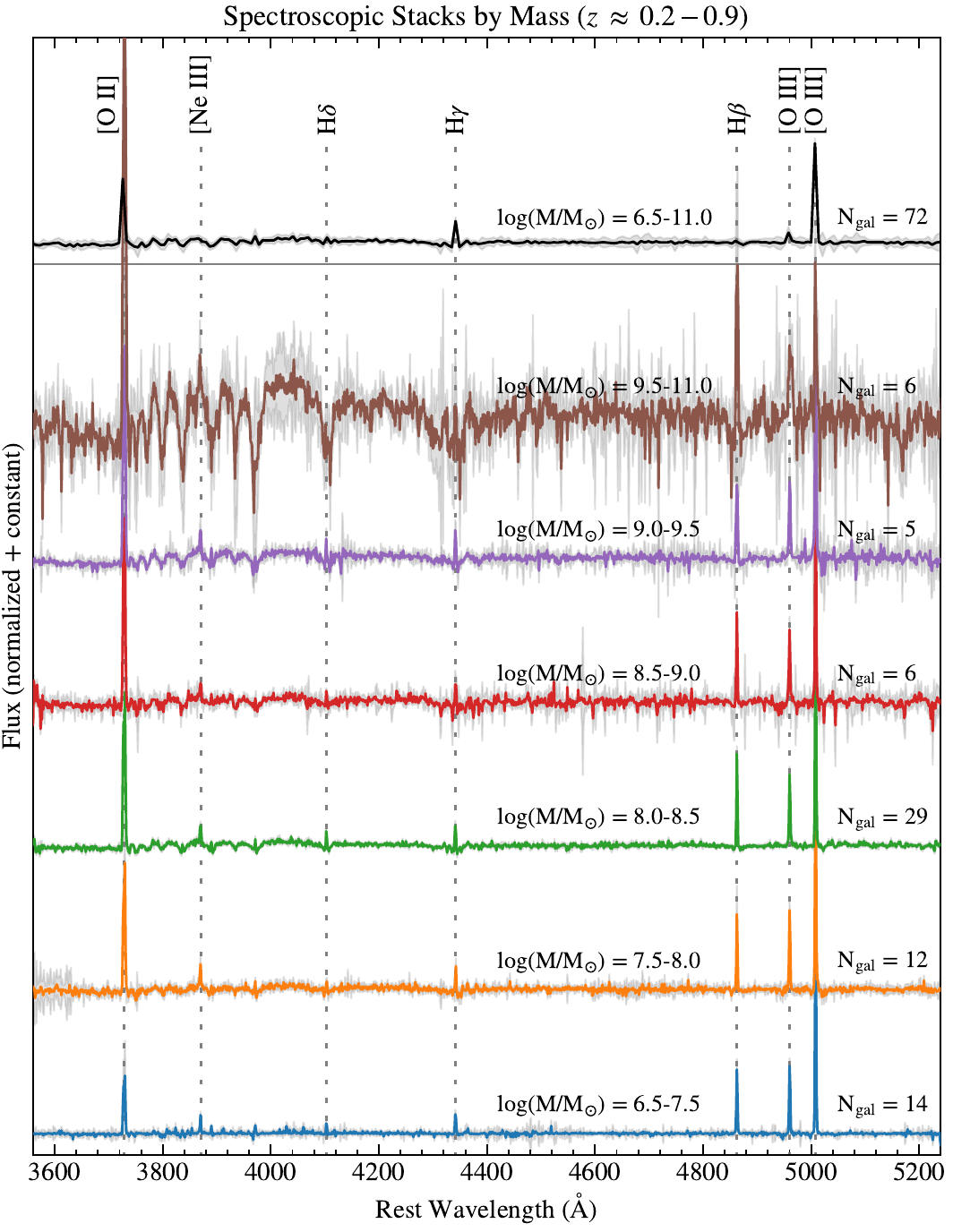}\\
\vspace{-0.8em}
\caption{The stacked and flux-normalized spectra for galaxies at $z\approx$~0.2 -- 0.9 in six bins of stellar mass with emission lines labeled. The total stack of all galaxies is shown in black, followed by stacks in order of decreasing stellar mass. The stacks are shown with different colors for clarity, with uncertainties represented by the gray shaded regions. The stellar mass range for each stack is shown above the spectrum and additional properties are given in Table~\ref{tab:results_stacks}. The fluxes at longer wavelengths are noisy due to the redshift distribution of targets over the adaptive optics notch filter wavelength gap ($\sim$5740~--~6000~\AA), which yields insufficient spectral coverage to determine robust fluxes in that spectral range. The monotonic variations in [O~III]/[O~II] and [O~III]/H$\alpha$ indicate notable changes in metallicity and ionization state with decreasing stellar mass.}
\label{fig:stacks_lowz}
\end{figure*}

\begin{figure*}[htb!]
\centering
\includegraphics[width=0.88\textwidth]{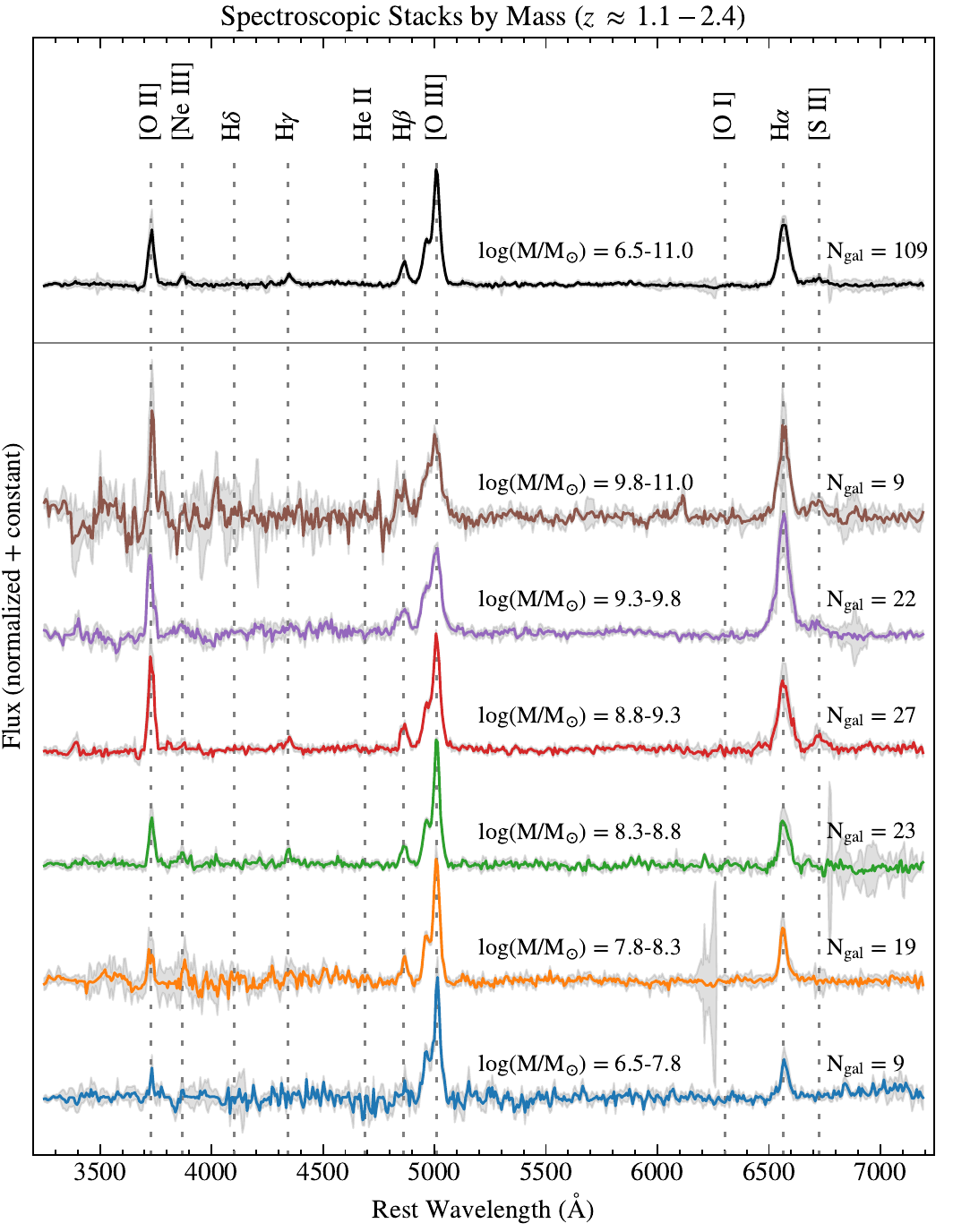}\\
\vspace{-0.8em}
\caption{The stacked and flux-normalized spectra for galaxies at $z\approx$~1.1 -- 2.4 in six bins of stellar mass with emission lines labeled. The total stack of all galaxies is shown in black, followed by stacks of decreasing stellar mass. Stacks are shown with different colors for clarity, with uncertainties represented by the gray shaded regions. The stellar mass range for each stack is shown above the spectrum with additional properties listed in Table~\ref{tab:results_stacks}. The increase of [O~III]/[O~II] and [O~III]/H$\alpha$ with decreasing stellar mass indicate changes in metallicity and ionization state.}
\label{fig:stacks}
\end{figure*}

We then determine the emission line fluxes using a Gaussian fitting procedure that is nearly identical to that described in \S\ref{ssec:fitting}, but that does not constrain the relative contribution of [N~II] to H$\alpha$ and also allows for multiple Gaussian components. Unlike \cite{Henry2021}, we find that we do not require both narrow and broad Gaussian components to fit our spectral stacks due to the smaller number of sources that result in overall narrower lines with lower S/N in the faint, extended wings. The spectral stacks for the low and high redshift samples are shown as functions of mass in Figures~\ref{fig:stacks_lowz} and \ref{fig:stacks}, respectively. The weak He~I $\lambda$5877 and $\lambda$6680 lines were also fit, but are not marked in the figures for clarity.

Our second goal is to characterize the metallicities of individual galaxies over all redshifts for comparison with the properties of their surrounding gas viewed in absorption along the quasar sightlines (Beckett et al. 2024, \textit{submitted}). While our spectral stacks only require a S/N~$\geq$~5 in the [O~III] emission line, sources near this limit have insufficient S/N to calculate metallicities individually. We therefore selected the 90 galaxies across all redshifts with S/N~$\geq$~10 in the [O~III] emission line for individual analysis. Using the strong [O~III] line for selection with this high S/N limit generally ensures detections of the metallicity sensitive lines while minimizing Eddington bias \citep{Eddington1913}. In order to avoid significant SFR selection biases, we confirmed that only requiring S/N~$\geq$~3 in the H$\beta$ emission line would produce a nearly identical sample, sharing 84 out of 90 of the same targets.

\subsection{Metallicity Methods}\label{ssec:calibrations}

The most abundant metal produced by stellar processes is oxygen, which serves as an excellent tracer of gas metallicity because it displays strong optical emission lines and is less depleted by dust than many other elements \citep{Stasinska2012}. The specific technique used to calculate the abundance of oxygen, and how that technique is calibrated, can substantially change the derived metallicity. The most common techniques for determining gas-phase metallicities are the direct method based on comparing temperature sensitive auroral emission lines with nebular lines from the same element and ionization state\footnote{As described in \cite{Strom2023}, auroral emission lines arise from forbidden transitions of electrons from the second to the first excited state, while nebular emission lines arise from the first excited to the ground state. The ratios of these lines can be used to calculate the electron temperature (T$_{e}$) in ionized gas at low densities where collisional de-excitation is negligible and the collisional excitation rate is primarily determined by the gas temperature.}, and the empirical strong-line method using bright nebular emission lines \citep{Maiolino2019}. The latter can be calibrated using photoionization models (e.g. \citealp{Kobulnicky2004}), or empirically from electron temperature diagnostics (e.g. \citealp{Pettini2004}).\footnote{The terms ``direct" and ``empirical" are misnomers, as all techniques require adopting model assumptions about the temperature and density distributions within the gas (see \citealp{Kewley2019}). While metal recombination lines can provide direct measurements of heavy element abundances, these lines are typically too weak to detect in extragalactic sources \citep{Esteban2014}.}

The direct method employs temperature sensitive line ratios to derive the electron temperature and corresponding emissivity for each ionization state of an element. The most frequently used ratios for deriving the abundances of O$^{+}$ and O$^{+2}$ are [O~II] $\lambda\lambda$7321,7332/$\lambda\lambda$3727,3730 and [O~III] $\lambda$4363/$\lambda$5007. An ionization correction is assumed for higher states that cannot be observed, but the amount of gas in the O$^{+3}$ and more highly ionized gas is negligible in star-forming galaxies with softer SEDs than AGN \citep{Berg2021}. The abundance in each ionization state is then summed to derive the total oxygen abundance. This technique has the advantage that the weak auroral lines remain strong at lower metallicities as oxygen becomes a more important avenue for cooling the gas. However, the auroral lines are typically $\sim$10~--~100 times weaker than the nebular lines they are compared to, so they can be difficult to detect observationally \citep{Yin2007}.

The strong-line method overcomes this limitation by only using the ratios of bright nebular emission lines in the spectra, but they must be empirically calibrated using sources where the direct method can also be applied, or by comparison with theoretical photoionization models. While this technique can be applied to large spectroscopic surveys with more limited S/N, this method has systematic uncertainties tied to the assumptions of the direct method or photoionization models used to calibrate the strong lines. Studies have found discrepancies as large as $\sim$0.7~dex between these two techniques when calibrating the strong-lines based on photoionization models \citep{Kewley2008, Stasinska2012}; however, strong-line diagnostics anchored by the direct method agree with recombination line metallicities at the $\sim$0.2~dex level \citep{Maiolino2019, Henry2021}. Critically, when multiple strong emission line ratios are employed, degeneracies between the metallicity, ionization parameter, gas density, and other properties are reduced, allowing for robust metallicity determinations \citep{Maiolino2019}.

\subsection{Metallicity Calculations}\label{ssec:calculations}

\subsubsection{Bayesian Method}\label{sssec:bayesian}

Considering the factors discussed in \S\ref{ssec:calibrations}, we adopt the methodology of \cite{Henry2021} that employs the direct method-based strong-line calibrations of \cite{Curti2017} to derive metallicites, and compare with the direct method for sources with auroral line detections. The justification for this choice of calibrations, their applicability at higher redshifts, and potential biases introduced by the emission line selection criteria are detailed comprehensively in \cite{Henry2021}. We also note that our adopted value of 12 + log(O/H)$_{\odot}$ = 8.69 $\pm$ 0.05 for the solar abundance of oxygen from \cite{Asplund2009} is identical to the value used in the \cite{Curti2017} calibrations, and matches other widely-used studies \citep{AllendePrieto2001} for an easier cross-comparison of results.

The Bayesian methodology developed by \cite{Henry2021} uses the $R_{2}\equiv$ log([O~II]/H$\beta$), $R_{3}\equiv$ log([O~III]/H$\beta$), and $O_{32}\equiv$ log([O~III]/[O~II]) strong-line calibrations presented in \cite{Curti2017} to calculate the most probabilistic metallicity. This process has the advantage that it self-consistently models the extinction from dust, stellar absorption of the Balmer lines, and contamination of H$\gamma$ by [O~III] $\lambda$4363 to marginalize over poorly-constrained parameters and derive realistic uncertainties on the metallicity estimates. First, in the redshift ranges with coverage of H$\alpha$ and H$\beta$, we use the observed ratio with a \cite{Calzetti2000} extinction curve to determine the dust extinction assuming an intrinsic Balmer decrement of H$\alpha$/H$\beta$ = 2.86. These measurements for each mass range can be used at adjacent redshifts without coverage of both lines to place appropriate priors on the extinction. The sample of \cite{Henry2021} included significantly more galaxies that are distributed randomly across the sky, and we therefore adopt identical extinction priors that have been matched to the closest mass bin for the high redshift sample.

\begin{figure}[b!]
\centering
\includegraphics[width=\columnwidth]{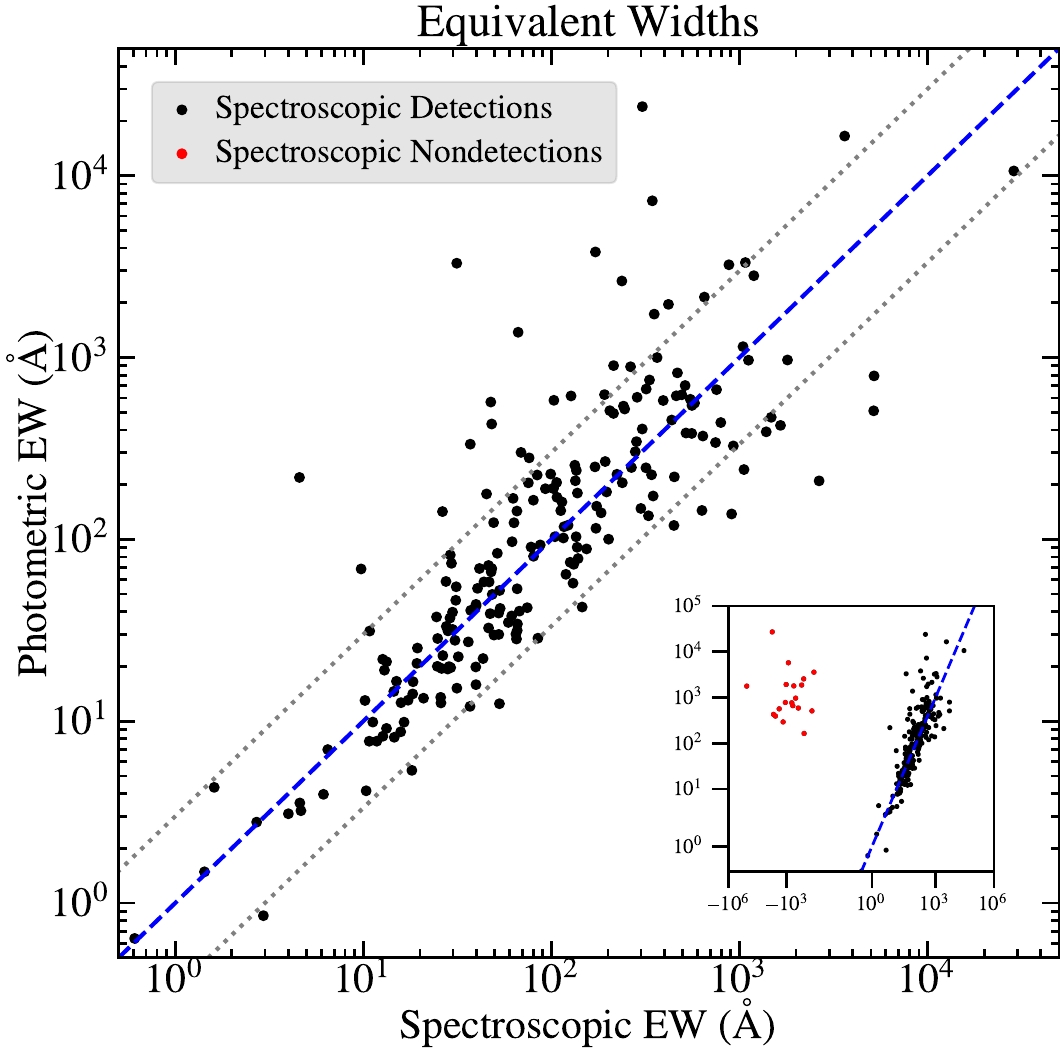}\\
\vspace{-0.5em}
\caption{A comparison of the equivalent widths (EWs) measured from the spectroscopic and imaging observations. The median systematic offset between the two measures is only 3\%. The main panel is shown on a logarithmic scale with a blue dashed unity line flanked by factor of three boundaries in dotted gray. The inset shows a larger parameter space using a symmetrical log (symlog) scale to highlight the sources without spectroscopic continuum detections (red points). While small changes in EW do not strongly affect the results, accurate order of magnitude estimates for all sources are important for deriving accurate metallicites. We adopt the broad-band photometric EWs to correct the Balmer emission line fluxes for stellar absorption.}
\label{fig:ews}
\vspace{-2em}
\end{figure}

Next, we account for Balmer emission line (H$\alpha$, H$\beta$, H$\gamma$, H$\delta$) stellar absorption by using a measure of the emission line equivalent width (EW) to reduce the model fluxes. The Bayesian framework allows for a range of stellar absorption based on both continuous and instantaneous burst models as described in Appendix~C of \cite{Henry2021}. \cite{Henry2021} also investigated measuring emission line EWs directly from WFC3 grism spectra as well as broad-band photometry. While estimates from the spectra are the most appropriate in principle, this process is sensitive to how the background is subtracted and fails for emission line sources where the continuum is not detected. We follow the procedure of \cite{Henry2021} and use our broad-band HST photometry to estimate EWs by subtracting line contamination and performing a linear interpolation across the filters to estimate the continuum flux at the wavelength of each emission line. At low redshifts we use the F336W and F702W for this purpose, while at higher redshifts we use the F125W and F140W photometry. We show a comparison of the spectroscopic and photometric EWs in Figure~\ref{fig:ews}, which demonstrates that the two estimates agree to within a systematic uncertainty of $\approx$~3\%, but the photometric technique is able to successfully derive EWs for all galaxies, including sources without spectroscopic continuum detections. The observed Balmer line fluxes are compared to the distribution of model values that have been reduced by the ratio of the emission line EW to the stellar absorption EW.

Finally, the emission line fluxes are compared with the \cite{Curti2017} calibrations to determine the most probabilistic value of the metallicity for each galaxy. This is completed using a Bayesian framework that considers extinction from dust, stellar absorption of the Balmer lines, contamination of H$\gamma$ by [O~III] $\lambda$4363, and the gas-phase metallicity using priors bounded over a reasonable range of parameter space. In most cases the priors have flat distributions except when informed by emission line diagnostics, with the extinction being a key example. \cite{Henry2021} explored the effects of correcting for reddening before or after spectral stacking and found that the derived metallicities agreed to within a few percent, indicating that the reddening treatment does not systematically bias the metallicities. The details of this process are documented in the appendices of \cite{Henry2021} and we refer to that study for a comprehensive explanation.

\subsubsection{Direct Method}\label{sssec:direct}

In order to confirm the validity of the abundances that we derive using the strong-line method, we use the direct method to calculate metallicities for galaxies with auroral line detections. We use our emission line catalog to identify six galaxies where the [O~III] $\lambda$4363 emission line is detected at S/N~$\geq$~4 (discarding IDs 1069 and 1023 due to blending), which is only possible using the MUSE data due to blending of [O~III] $\lambda$4363 and H$\gamma$ in the WFC3 spectra. In addition, we select six galaxies where the [O~III] $\lambda$1666 emission line is detected at S/N~$\geq$~4 in MUSE. While this line is rarely detected beyond the local Universe except in gravitationally lensed sources \citep{Sanders2020, Citro2023}, it provides a similar temperature diagnostic to [O~III] $\lambda$4363 in the rest-frame UV \citep{Mingozzi2022}, which is covered by MUSE at $z \approx$~2.

We visually inspect each source and remove those with S/N~$\leq$~5 that display noisy line profiles. We then use PyNeb (v1.1.17; \citealp{Luridiana2015}) to convert the [O~III] $\lambda$4363 / $\lambda$5007 and $\lambda$1666 / $\lambda$5007 ratios to electron temperatures, and calculate the ionic abundances of [O~III] and [O~II]. We measure the [O~III] gas temperature directly with the auroral lines, and use the T$_e$(O$^{+}$)~--~T$_e$(O$^{++}$) relation of \cite{Campbell1986} to derive the [O~II] gas temperature. We assume an electron density of n$_e =$~250~cm$^{-3}$, which is typical of star-forming galaxies at these redshifts \citep{Sanders2018}. We note that the metallicities are robust across a wide range of densities from n$_e =$~100~$-$1000~cm$^{-3}$ \citep{Steidel2014}.

We adopt a Galactic extinction curve \citep{Savage1979} to correct the line ratios for reddening using the H$\gamma$/H$\beta$ and H$\alpha$/H$\beta$ ratios for the low redshift sources and conservatively adopt the lower of the two extinctions. At high redshifts, we only have wavelength coverage of the H$\gamma$/H$\beta$ ratio, which is contaminated by [O~III] $\lambda$4363. We use the observed H$\gamma$/H$\beta$ ratios to determine a lower limit on the reddening, as well as assuming H$\gamma$ is contaminated by 20\% from [O~III] $\lambda$4363 for an upper limit on the reddening. The later only lowers the metallicity by $>$~0.1~dex for one source, and so we correct the line ratios using the observed H$\gamma$/H$\beta$ ratios.

Finally, the uncertainties for our strong-line metallicities are propagated from the Bayesian analysis, while those for the direct method are the quadrature sum of the uncertainties in the emission line fluxes that are used in the line ratios for the direct method. Specifically, [O~III] $\lambda$4363 / $\lambda$5007 and $\lambda$1666 / $\lambda$5007 used to derive electron temperatures, as well as [O~III]/H$\beta$ and [O~II]/H$\beta$ that are used to determine the ionic fractions. Considering the weakness of the auroral lines, and the need to measure fluxes for [O~II], [O~III], and H$\beta$, the comparison of these techniques is only possible for six sources at $z \approx$~0.68~--~0.78 using [O~III] $\lambda$4363, and for six additional sources at $z \approx$~2.25~--~2.32 using [O~III] $\lambda$1666. The spectra for these 12 sources are shown in the \hyperref[app:spectra]{Appendix}.

\section{Results}\label{sec:results}

\subsection{Validating Strong-line Calibrations with the \texorpdfstring{T$_{e}$}{Te} Method}\label{ssec:direct}

\begin{figure}[ht]
\centering
\includegraphics[width=\columnwidth]{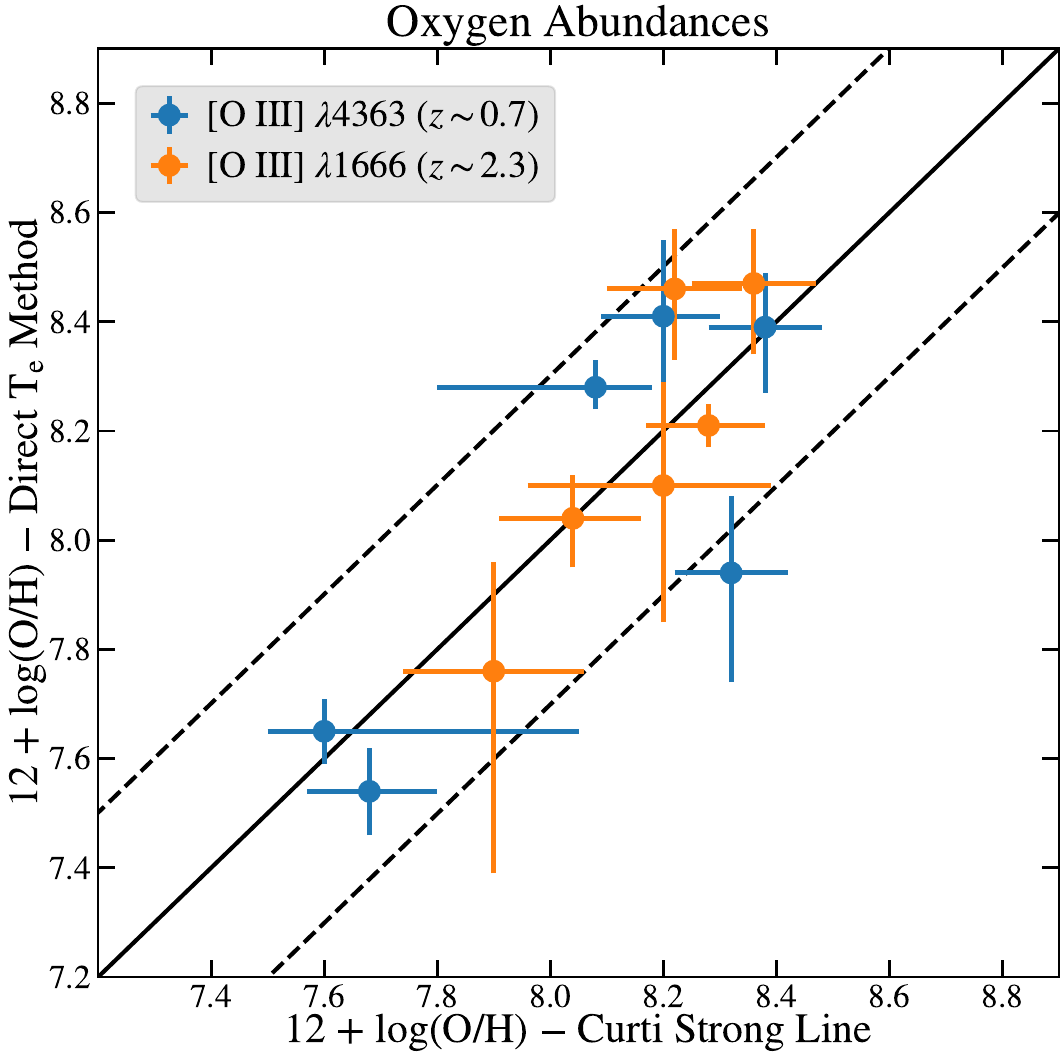}
\caption{A comparison of the gas-phase oxygen abundances derived using our Bayesian framework with the \cite{Curti2017} strong-line calibrations, versus those found using the direct T$_\mathrm{e}$ method with PyNeb. The solid unity line is flanked by factor of two dashed boundaries ($\pm$~0.30~dex). The results from these techniques agree to within a factor of two within the uncertainties for all sources, with the lowest metallicity sources at log(O/H) + 12 $\approx$~7.7 having robust [O~III] $\lambda$4363 detections at S/N~$>$~10. The results for [O~III] $\lambda$1666 validate the use of these strong-line calibrations up to $z \approx$~2.5. The data shown in this figure are available in the Appendix in Table~\ref{tab:direct}.}
\label{fig:direct}
\end{figure}

In Figure~\ref{fig:direct}, we compare the metallicities derived using our strong-line methodology based on the \cite{Curti2017} calibrations to the metallicities determined from the direct T$_\mathrm{e}$ method using PyNeb. The blue circles represent [O~III] $\lambda$4363 measurements at $z \approx$~0.7, while the orange circles are [O~III] $\lambda$1666 measurements at $z \approx$~2.3. Overall, there is excellent agreement for all of the sources at both low and high redshift with a systematic offset of only 0.03~dex and a dispersion of 0.17~dex between the two methods. Importantly, this agreement is seen for the lowest metallicity sources where the [O~III] $\lambda$4363 line is detected at S/N~$>$~10. The remaining [O~III] $\lambda$4363 sources have weaker detections that may drive the scatter. The comparison of the six sources at $z \approx$~2.3 represents a fundamental result, as there are only a few [O~III] $\lambda$1666 metallicity determinations at these redshifts for unlensed sources \citep{Sanders2020}. Recently, \cite{Llerena2023} reported 21 [O~III] $\lambda$1666 detections at $z \approx$~3, with just five having S/N~$\geq$~5. These detections required VUDS and VANDELS spectroscopy with $\sim$20~$-$~80 hour integration times, highlighting the difficulty of detecting this auroral line at high redshift even with very deep observations. We approximately double existing samples at $z >$~1 with detections at S/N~$\geq$~5 for unlensed systems, and the results show the validity of using these strong-line calibrations up to $z \approx$~2.5.

The fact that the strong-line and direct-method results agree for the lower redshift [O~III] $\lambda$4363 sources where all of the emission lines are in MUSE, as well as for the [O~III] $\lambda$1666 sources where [O~III] $\lambda$5007 is in the HST grism, gives confidence in the inter-flux calibration between the instruments. Finally, the derived extinction for these sources is low, and consistent with zero within the uncertainties in several cases, which is expected for low-mass sources and mitigates concerns about the choice of reddening correction for the widely spaced [O~III] $\lambda$5007 and $\lambda$1666 emission lines.

\subsection{The Mass-Metallicity Relation}\label{ssec:mzr}

\begin{figure*}[htb!]
\centering
\includegraphics[width=\textwidth]{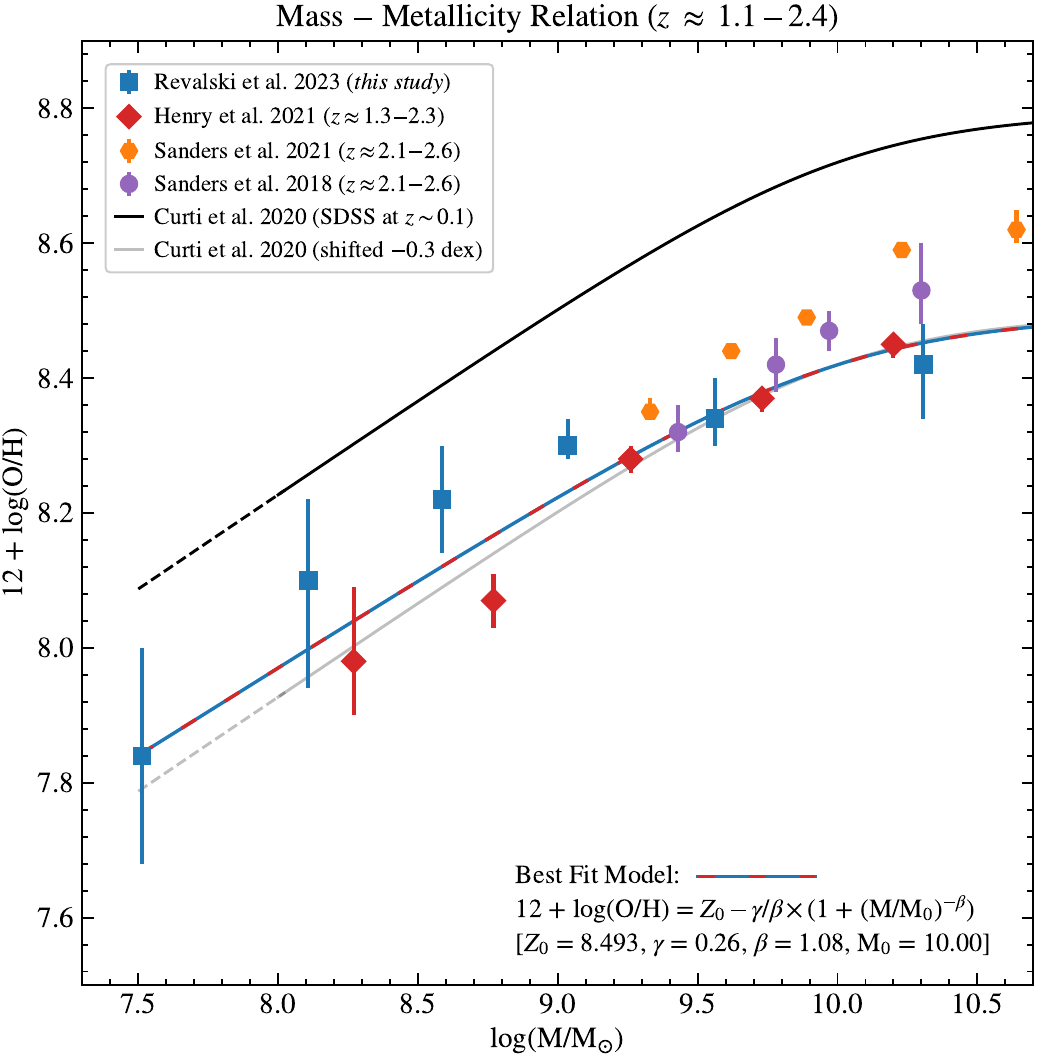}\\
\vspace{-0.5em}
\caption{The mass-metallicity relation (MZR) for galaxies at $z\approx$~1.1 -- 2.4, with the results for our stacked spectra shown with blue squares. The MZR decreases to log(O/H) + 12 $\approx$~7.8 $\pm$ 0.1 (15\% solar) at log(M$_{\star}$/M$_{\odot}$)~$\approx$~7.5, without any evidence of a turnover or flattening in the shape of the MZR at low stellar masses. The overall shape of the MZR is consistent with the local relations of \cite{Andrews2013} and \cite{Curti2020} when these relations are shifted by $\sim$0.3~dex to lower metallicities as shown by the gray line. This shift is primarily driven by galaxies having lower metallicities at higher redshifts, but also has a secondary dependence on SFR. As compared with \cite{Henry2021}, our stacks reach $>$~0.5~dex lower in SFR so the trending of our results to higher metallicities is expected and consistent with the M-Z-SFR relation. We have used published emission line fluxes to calibrate the results of \cite{Sanders2018, Sanders2021} to the \cite{Curti2017} relation for equal comparison across studies. We use the same methodology as \cite{Henry2021} and thus derive a best fit model for our combined results shown with the dashed blue-red line. The equation from \cite{Curti2020} and our best fit model parameters are shown in the lower-right of the figure.}
\label{fig:results_stacks}
\end{figure*}

We present the gas-phase mass-metallicity relation (MZR) at $z \approx\,$1$\,$--$\,$2 for our sample in Figure~\ref{fig:results_stacks}. The stacked spectra reach stellar masses of M$_{\star}$~$\approx$~10$^{7.5}$~M$_{\odot}$ at SFRs of $\sim$~0.3~M$_{\odot}$ yr$^{-1}$, which is approximately six times lower in stellar mass, and an order of magnitude lower in SFR, than earlier studies using HST in this redshift range \citep{Sanders2018, Sanders2021, Henry2021}. This is also $\sim$1.5~dex lower in mass than the majority of previous studies that were limited to log(M$_{\star}$/M$_{\odot}$)~$>$~9.0 at these redshifts before JWST. We find that the MZR decreases to log(O/H) + 12 $\approx$~7.8 $\pm$ 0.1 (15\% solar) at log(M$_{\star}$/M$_{\odot}$)~$\approx$~7.5, without evidence of a turnover or flattening in the shape of the MZR at the lowest masses. The shape of the MZR across all masses is consistent with the local relations derived for SDSS galaxies \citep{Andrews2013, Curti2020}, except shifted to lower metallicities. We confirm that the MZR resides $\sim$0.3~dex lower in metallicity than local galaxies, but note that the magnitude of the offset is a function of both metallicity and SFR, as discussed in \S\ref{ssec:fmr}.

In Figure~\ref{fig:results_stacks} we also display the results of \cite{Henry2021} and \cite{Sanders2018, Sanders2021} at similar redshifts. We have used published emission line fluxes to recalibrate the \cite{Sanders2018, Sanders2021} results to the \cite{Curti2017} calibration for an equal comparison across studies. Overall, we see excellent agreement between the studies at high masses, with some deviations between our result and that of \cite{Henry2021} at lower masses, which are primarily due to the different average SFRs of the samples. These results are also very consistent with the recent study by \cite{He2024} at similar redshifts.

\begin{figure}[b!]
\centering
\includegraphics[width=\columnwidth]{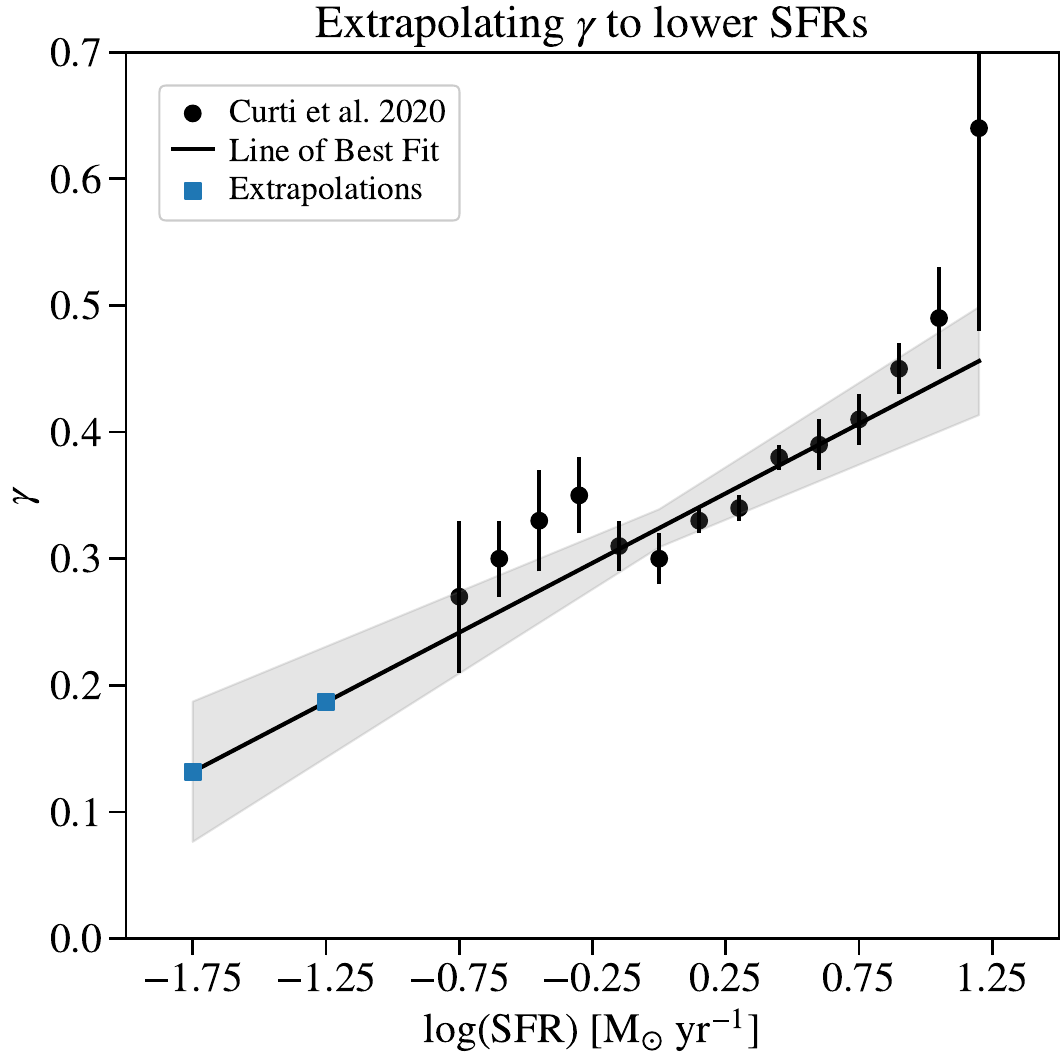}
\caption{The dependence of the low-mass slope ($\gamma$) in Equation~\ref{eqn:mzr} on the SFR of the galaxy. The values and uncertainties from Table~5 of \cite{Curti2020} are shown with black circles, while the extrapolated values for log(SFR) = $-$1.25 M$_\odot$ yr$^{-1}$ ($\gamma$ = 0.18) and $-$1.75 M$_\odot$ yr$^{-1}$ ($\gamma$ = 0.13) are shown as blue squares. The shaded region represents the error-weighted uncertainty on the slope (0.11) and intercept (0.32) for the line of best fit, shown by the solid back line.}
\label{fig:gamma}
\end{figure}

\begin{figure*}[htb!]
\centering
\includegraphics[width=0.49\textwidth]{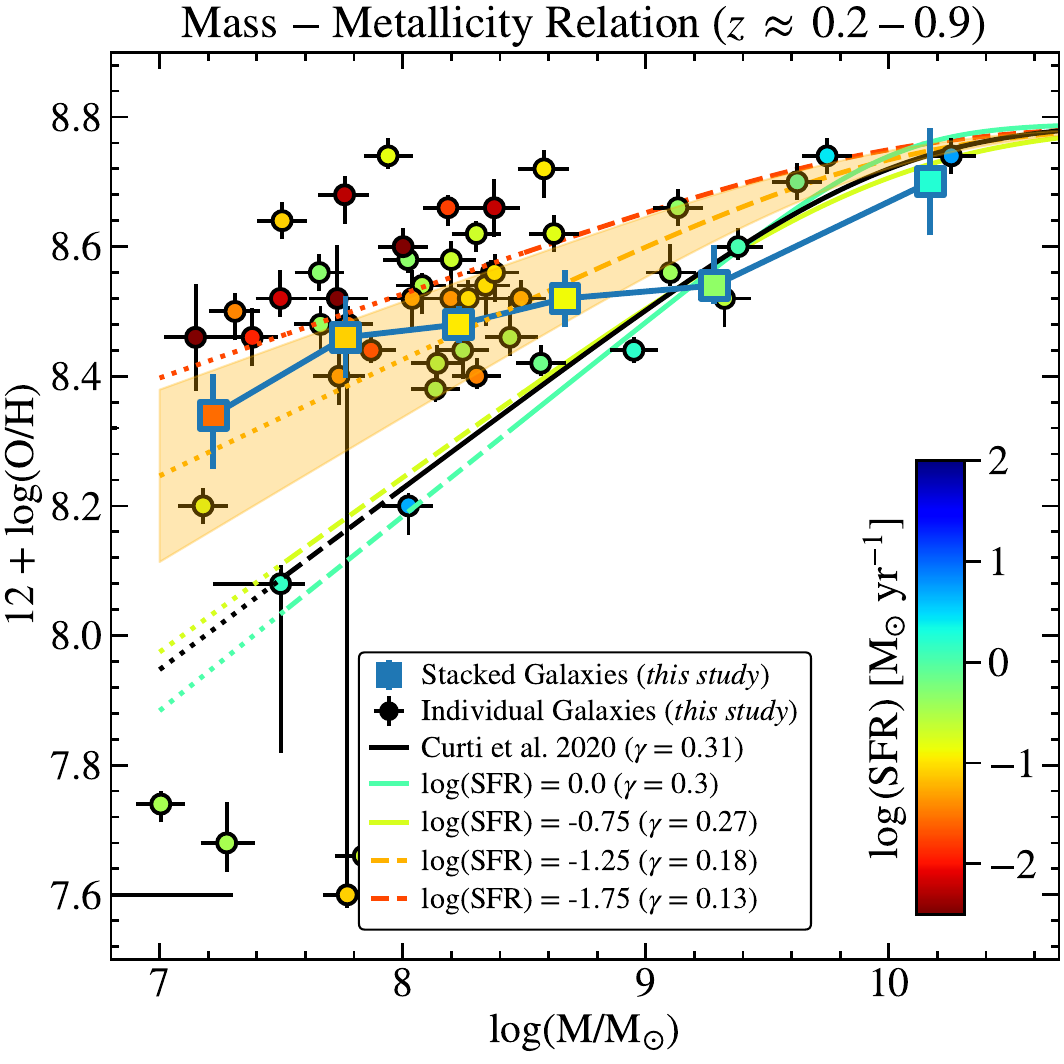}
\includegraphics[width=0.49\textwidth]{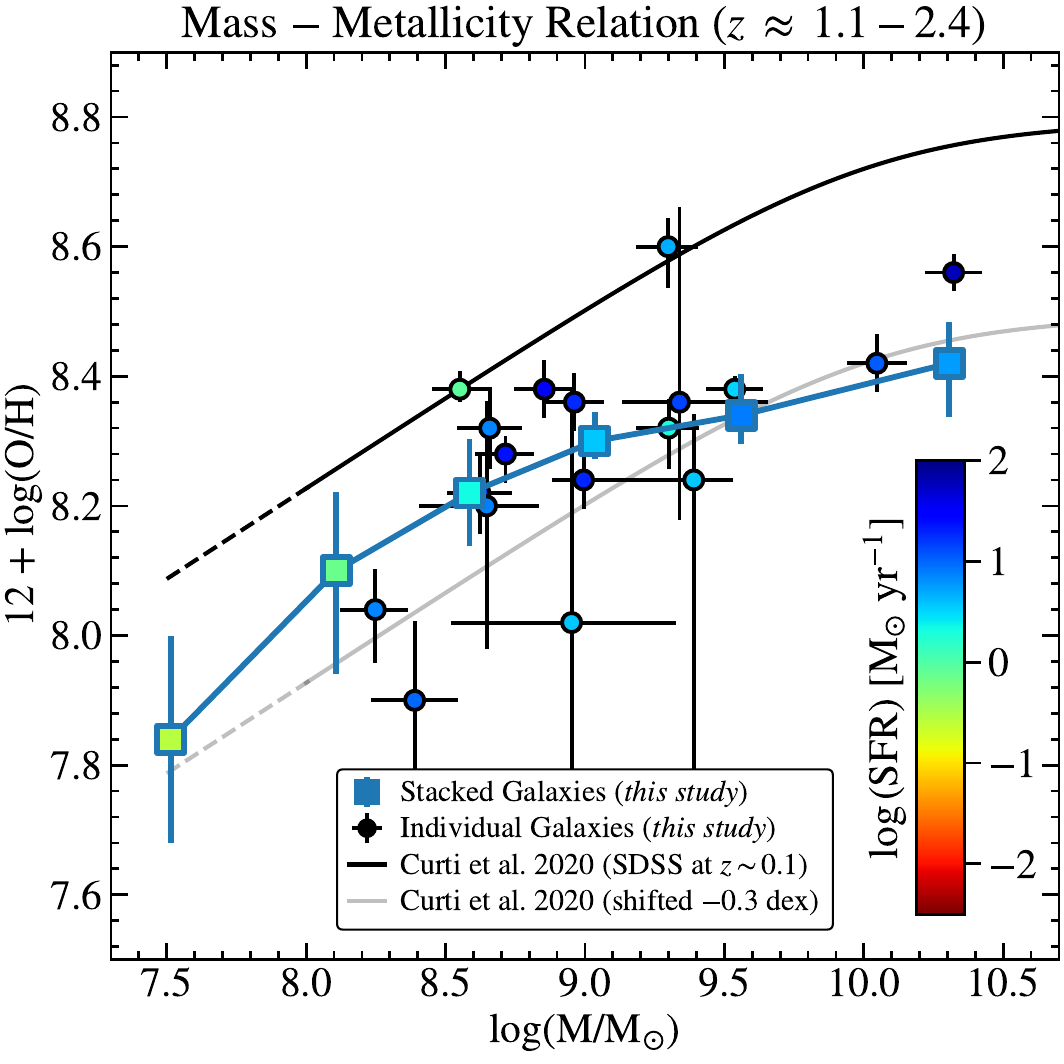}
\caption{The mass-metallicity relation (MZR) for galaxies at $z\approx$~0.2 -- 0.9 (left) and $z\approx$~1.1 -- 2.4 (right). The results for stacked spectra are shown with joined blue squares color-coded by their median SFRs, while the points for individual galaxies with [O~III] S/N $\geq$ 10 are shown as circles color-coded by their model-based SFRs. In the left panel, we show the local MZR from \cite{Curti2020} with a black line, with values for specific SFRs color-coded. The solid sections of each line are from \cite{Curti2020}, while the dashed sections are extrapolations to lower masses or higher SFRs, and the dotted sections are extrapolated across both mass and SFR. The shaded region shows the uncertainty in the extrapolation of $\gamma$ = 0.18 for log(SFR) = $-$1.25 M$_\odot$ yr$^{-1}$. In the right panel, our results are consistent with the local MZR after it is shifted by $\sim$0.3~dex to lower metallicities to account for redshift evolution and different SFRs. The vertical offset from the local MZR in the left panel is driven by the lower SFRs of the stacks as compared to SDSS galaxies, while the offset in the right panel is primarily due to redshift evolution.}
\label{fig:results_individuals}
\end{figure*}

We used the \href{https://lmfit.github.io//lmfit-py/}{\textsc{lmfit}} package \citep{Newville2023} to simultaneously fit our results with those of \cite{Henry2021}, as we employ equivalent methodologies spanning a large range in SFR. We fit the MZR using the same approach as \cite{Curti2020}, who characterize the MZR using a functional form similar to a Lomax (or Pareto Type II) distribution. Specifically,
\begin{equation}
12 + \mathrm{log(O/H)} = Z_0 - \gamma / \beta \times (1 + (M/M_{0})^{-\beta}),\label{eqn:mzr}
\end{equation}
\noindent
where $Z_0$ is the maximum metallicity, $M_{0}$ is the turnover mass, $\gamma$ is the low mass slope, and $\beta$ is the width of the turnover. In general, $M_{0}$, $\gamma$, and $\beta$  are functions of SFR. The best-fit line and the corresponding parameter values are shown in Figure~\ref{fig:results_stacks}, which are nearly identical to that of \cite{Curti2020} after shifting it by 0.3~dex to lower metallicities to account for redshift evolution.

\subsection{The Mass-Metallicity-SFR Relation}\label{ssec:fmr}

It can be misleading to compare the MZRs from different studies unless they are derived for galaxies with similar SFRs. As discussed in \S\ref{sec:intro}, galaxies of a specific mass show lower metallicities at higher SFRs (e.g. see Figures 6 and 9 in \citealp{Curti2020}). This dependence on SFR defines a three-dimensional plane of Mass-Metallicity-SFR (M-Z-SFR) that is often termed the Fundamental Metallicity Relation (FMR; \citealp{Mannucci2010}). We can account for this dependence on SFR using two different approaches. Commonly in the literature, the low mass slope ($\gamma$) is fixed and the 3D plane can be reduced to a 2D projection of least scatter by scaling the stellar masses using a fraction of the SFR, which is parameterized by $\mu$ = log(M$_{\star}$) -- $\alpha$log(SFR), where $\alpha$ quantifies the degree to which metallicity and SFR are correlated. Studies have found various values for $\alpha$ due to differences in galaxy selection criteria and metallicity calibrations \citep{Andrews2013, Curti2020}. This projection effectively aligns MZRs with different SFRs to a common location on the MZR diagram. In the case of SDSS galaxies, the range of $\gamma$ values is sufficiently modest that this approach removes nearly all of the spread in the MZR introduced by SFR.

However, at low redshifts our results reside well above the local MZR relations of \cite{Andrews2013} and \cite{Curti2020}. This offset is expected as our sample reaches much lower SFRs than the average for SDSS galaxies. Thus, we use an alternative approach and account for the correlation between the low mass slope of the MZR ($\gamma$) and the SFR by accounting for $\gamma$ as a function of SFR. While $\gamma$ is now a variable parameter, this approach eliminates the need to determine a projection of least scatter, which can convolve the uncertainties in mass and SFR in a complex manner. As shown in Figure~\ref{fig:gamma}, we use the values provided in Table~5 of \cite{Curti2020} to extrapolate $\gamma$ to the lower SFRs covered by our observations, with extrapolations shown for log(SFR) = $-$1.25 ($\gamma$ = 0.18) and $-$1.75 ($\gamma$ = 0.13) M$_\odot$ yr$^{-1}$.

We show the M-Z-SFR relation for our low and high redshift samples in Figure~\ref{fig:results_individuals}, with the results for individual galaxies color-coded based on their SFRs and the extrapolated relations shown for the low redshift sample. The uncertainties in the masses are derived from the SPS models added in quadrature with a $\pm$0.1 dex systematic model uncertainty, while those in the abundances are from the Bayesian model added in quadrature with a $\pm$0.02 dex uncertainty from the model grid resolution. Overall, we find excellent agreement between our results and the extrapolations, which agree within the uncertainties to better than one standard deviation. These results indicate that the local MZR is valid to at least $\sim$1~dex lower in SFR than current calibrations based on SDSS.

At both low and high redshifts the results for the individual galaxies are consistent with those derived for the stacked spectra, with measurements scattered above and below the stacked results. This indicates that the stacks are not biased in metallicity as compared to the individual galaxies. However, the stacks do reach lower SFRs than measurements for the individual sources at high redshift due to S/N limitations. There is also no noticeable trend in SFR for sources above and below the stacked results within a given mass bin, suggesting that the sample is complete over the ranges where both individual and stacked metallicities can be derived. However, there are an insufficient number of sources in the high redshift sample to detect such a trend if it is weakly present. The results shown in Figures~\ref{fig:results_stacks} and \ref{fig:results_individuals} are tabulated in Tables~\ref{tab:results_stacks} and \ref{tab:results_individuals}, respectively, for easier comparison with other MZR studies.

As seen in Figure~\ref{fig:results_individuals}, our low redshift sample resides well above the local MZR. This result is expected, as the SDSS calibrations adopt a constant slope ($\gamma$ = 0.31) for the low mass end of the relation, which neglects the correlation between the low mass slope and the SFR. This is appropriate for the range of masses, metallicities, and SFRs of local SDSS galaxies, and also safely prevents extrapolation errors. However, our deep observations reach log(SFR)~$\approx$~$-$2.5, which is significantly lower than where the existing calibration is constrained. Without accounting for this dependence of $\gamma$ on the SFR, galaxies can artificially appear to reside above or below the FMR. These results indicate that expanding calibrations to larger samples at lower masses and SFRs with JWST and Roman may require incorporating the correlation between $\gamma$ and SFR, which avoids correlated errors between mass and SFR that occur when using the projection of least scatter.

\begin{figure*}[htb!]
\centering
\includegraphics[width=0.49\textwidth]{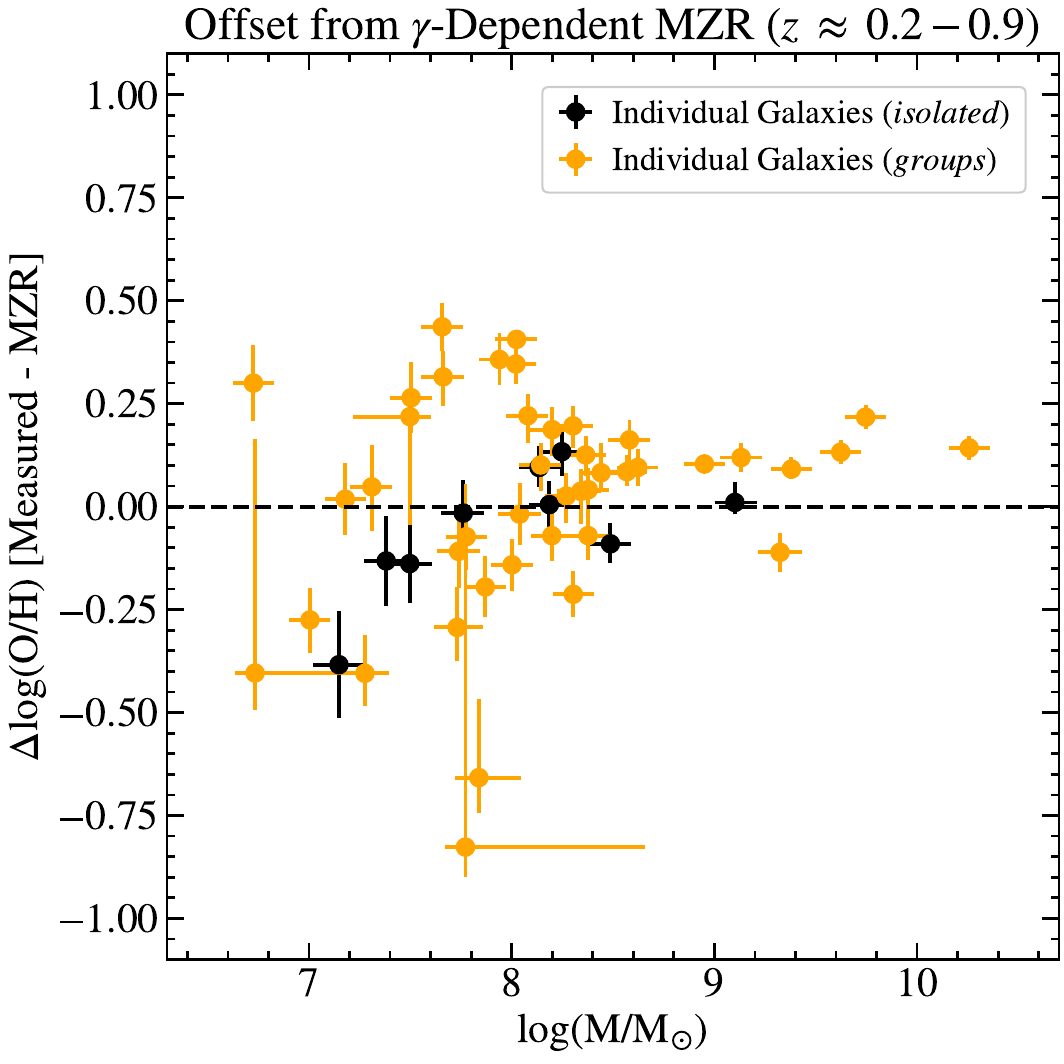}
\includegraphics[width=0.49\textwidth]{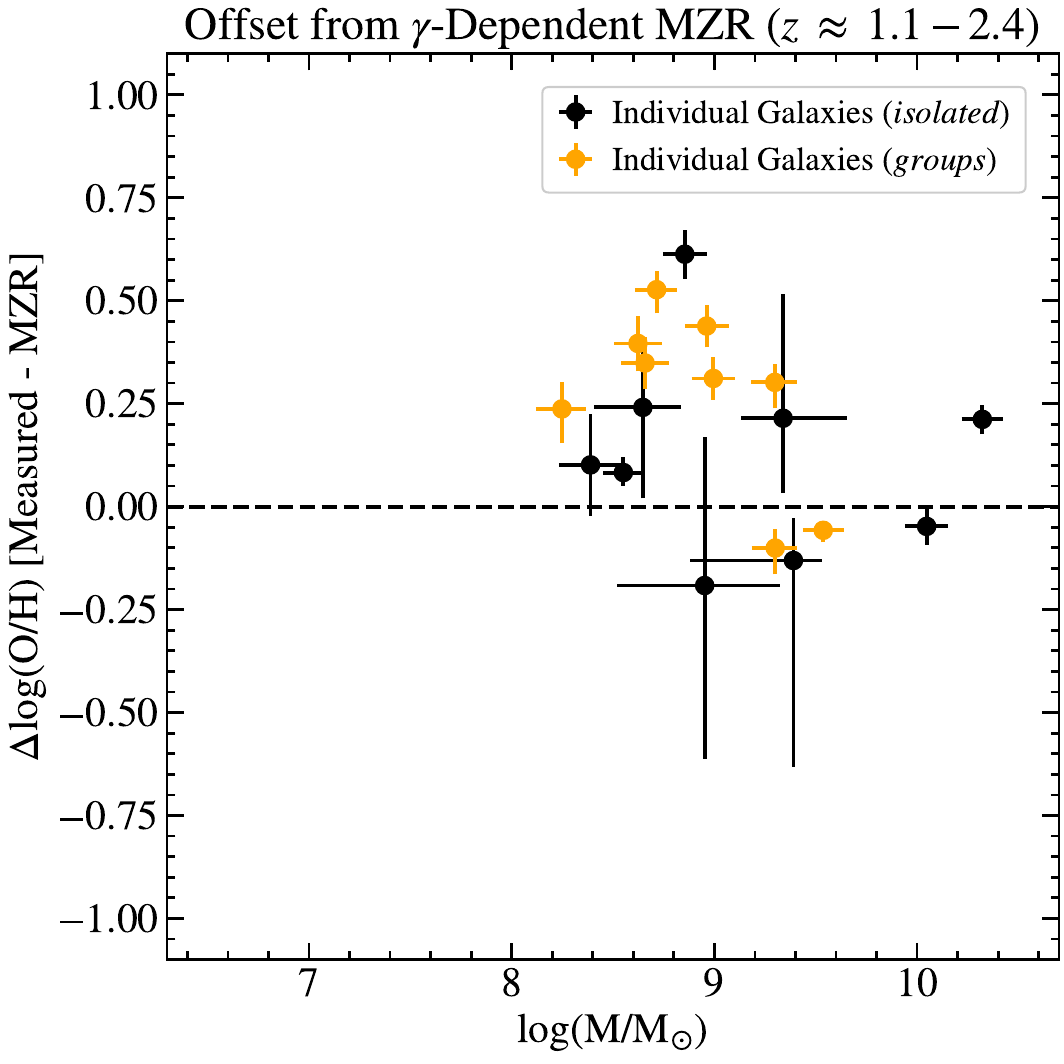}
\caption{The offset from the MZR for individual galaxies at $z\approx$~0.2 -- 0.9 (left) and $z\approx$~1.1 -- 2.4 (right) using the best-fit $\gamma$-values that were extrapolated based on the SFR of each source (see Figure~\ref{fig:gamma}). The results for sources with [O~III] S/N $\geq$ 10 are color-coded based on whether they are isolated galaxies (black), or reside in a group environment (gold). The best-fit $\gamma$ values encompass $\gamma$ = 0.1 -- 0.3 for the low redshift sample, and $\gamma$ = 0.3 -- 0.4 at high redshifts. The small dispersion of sources centered on zero indicates that the $\gamma$-dependent MZR accurately encapsulates the large range of observed SFRs. Overall, the offset from the MZR for galaxies in groups are the same as isolated galaxies to within 0.1~dex (left) and 0.2~dex (right) in metallicity. The errors are propagated from those in the metallicities and the uncertainty in extrapolating $\gamma$ to lower SFRs.}
\label{fig:results_groups}
\end{figure*}

\subsection{The Role of Environment}\label{ssec:environment}

Finally, we investigate the influence of galaxy environment on the metallicities of individual galaxies. We present the differences between the measured metallicities for individual sources and the values predicted by the MZR using our best fit value of $\gamma$ for each galaxy in order to control for SFR in Figure~\ref{fig:results_groups}, with the points color-coded based on whether the galaxies reside in isolated (black) or group (gold) environments. We used our group catalog developed in \cite{Dutta2023} to separate galaxies by environment, where the catalog was obtained by running a Friends-of-Friends (FoF) algorithm on the galaxy sample with reliable redshifts using linking lengths of 400~kpc in the transverse spatial direction and 400 km~s$^{-1}$ in the line-of-sight velocity direction.

In the low redshift sample, we find that the majority of galaxies are in group environments, which is expected given the presence of a large group at $z \approx$~0.68 \citep{Fossati2019}. In the high redshift sample, half of the sources with individual metallicities are in group environments located at $z \approx$~2.25 and $z \approx$~2.38. We calculated the mean and median metallicities for the isolated and group galaxies and find that the metallicities of galaxies in group environments are equal to isolated sources to within 0.1~dex for the low redshift sample, and 0.2~dex for the high redshift sample. However, this slight enhancement of metallicity in group galaxies is not statistically significant according to the KS test, due to the small sample sizes and scatter in the individual metallicity measurements. The relatively small scatter surrounding zero indicates that the variable $\gamma$ approach accurately captures the location of each galaxy on the M-Z-SFR plane.

\section{Discussion}\label{sec:discussion}

The successful launch of JWST \citep{Gardner2006, Gardner2023} provides an opportunity to study galaxies at lower masses and SFRs and higher redshifts than previously possible. This creates a timely need to validate existing metallicity calibrations, and to extend them over larger ranges of physical parameters. Our results using the direct method indicate that the existing strong-line calibrations from \cite{Curti2020} are valid up to at least $z \approx$~2.5 across a broad range of masses and SFRs.

However, recent studies with JWST suggest that common strong-line calibrations may not be valid at $z >$~3. Specifically, \cite{Laseter2023} compared the metallicities of 10 sources at $z \approx$~2~$-$~9.5 and found that strong-line calibrations failed to match direct method measurements. \cite{Curti2023} expanded this investigation to 146 star-forming galaxies and found evidence that the slope of the MZR may begin to flatten at lower masses, and that galaxies increasingly diverge from the FMR beyond $z \approx$~3. Evidence of a redshift evolution in the MZR and FMR and a lack of consistency of local strong-line calibrations with the direct method are supported by several recent JWST studies \citep{Li2023, Nakajima2023}.

The results of these studies have important implications for the evolution of the MZR and FMR with mass, SFR, and redshift. Historically, the low mass regime of the MZR has been represented as a decreasing power-law. However, the relationship must turnover at some critical mass and approach a metallicity floor, otherwise the lowest mass galaxies would consist of only hydrogen and helium. Studies have been searching for this flattening of the MZR, with evidence at the lowest masses observed for local galaxies \citep{Hirschauer2022}. The results of \cite{Curti2023} and others suggest that we are beginning to observe a flattening in the low-mass slope of the MZR across a range of redshifts at $z >$~3.

Interestingly, we do not see evidence of a flattening or turnover in our results (Figure~\ref{fig:results_stacks}). The overall shape is very consistent with local MZR relations after they are shifted by $\sim$0.3~dex to lower metallicities to account for redshift evolution. The results for our lowest mass stack in the high redshift sample reside almost exactly between the \cite{Andrews2013} and \cite{Curti2020} relations that have been extrapolated to log(M$_{\star}$/M$_{\odot}$)~$\approx$~7.5. The lowest mass stack also has the lowest SFR, suggesting that galaxies with higher SFRs at this mass would have even lower metallicity. Nonetheless, we do not see evidence for a flattening or turnover in the MZR at these masses, SFRs, and redshifts. Given the importance of our lowest mass stack in the high redshift sample, we further investigated the effects of having a small number of targets within the stack. In addition to the boot-strapped uncertainties, we confirmed the robustness of this measurement by adding one, two, and three of the next highest mass targets into the stack. While the average mass of the bin shifted minutely, the derived metallicities agree to within 0.04~dex, which is the size of our Bayesian grid interval, in all three cases.

However, assuming constant $\gamma$, we see evidence that our results deviate from the local MZR. We observe that our low redshift sample resides above it by $\sim$0.2~dex, while the high redshift results are consistent with it. In these cases, the assumption that the low mass slope ($\gamma$) is constant with SFR is no longer valid. Accounting for the low mass slope as a function of SFR brings our results into agreement with the local MZR, without including any direct redshift evolution. 

These results provide a crucial constraint on the MZRs and FMRs at low stellar masses, a regime in which modern hydrodynamical simulations are currently limited. Simulation suites such as SIMBA \citep{Dave2019}, IllustrisTNG \citep{Vogelsberger2014, Genel2014, Springel2018}, FIRE \citep{Hopkins2014, Hopkins2018, Muratov2015}, and EAGLE \citep{Schaye2015} require a compromise between the cosmological volume that is probed and mass resolution, such that low mass systems can only be studied in smaller volumes. With advancements in computation, the MZR and FMR have now been extended below $\sim$10$^{8.5}$M$_\odot$ \citep[e.g.,][]{Ma2016, Torrey2018, Torrey2019, Dave2019, Feldmann2023}. These works have successfully reproduced the broad slope and normalization of these relations across a range of redshifts; however, the low-mass end of the MZR and FMR require further investigation in observations, simulations, and models \citep{Ucci2023}. Hence, this work can serve as benchmark for future observations and simulations.

As the archive of deep, high-redshift observations with JWST grows, and the quality of the data calibrations steadily improve, emission line calibrations are being expanded to a vastly increased parameter space in a robust, self-consistent manner. Progress is already being made in this area with revised calibrations proposed by \cite{Sanders2023}, \cite{Hirschmann2023}, and additional studies \citep{Nakajima2022, Garg2023, Yang2023}. This rapid release of results related to the metallicities of galaxies across cosmic time should provide a consensus on these issues that was not possible before JWST (e.g. \citealp{Atek2023, Bagley2023, Curti2023b, Heintz2023, Langeroodi2022, Langeroodi2023, Maseda2023, Matthee2023, Shapley2023, Sun2023, Trump2023}). These investigations are crucial, as our results approach the low metallicity limit of the current strong-line calibrations. Critically, our direct method metallicities for the [O~III] $\lambda$1666 emitters validates the use of the existing \cite{Curti2017} strong-line calibrations up to $z \approx$~2.5.

\newpage
\section{Conclusions}\label{sec:conclusions}

We have used ultra-deep imaging and spectroscopy of the MUSE Ultra Deep Field to characterize the gas-phase metallicities of low mass galaxies up to $z \approx$~2.3 during the peak of cosmic star-formation. Our conclusions are the following:

\begin{enumerate}
\item We find excellent agreement between the metallicities derived from the \cite{Curti2017} strong-line calibrations and the direct method for the 12 sources with detections of the [O~III] $\lambda$4363 ($z\approx$~0.7) and [O~III] $\lambda$1666 ($z\approx$~2.3) auroral lines. The results of these techniques agree to within a factor of two (0.3 dex) within the uncertainties for all galaxies, which validates the use of these strong-line calibrations up to $z\approx$~2.5. The six [O~III] $\lambda$1666 detections with S/N~$>$~5 nearly doubles the number previously reported at $z >$~2 for unlensed galaxies, providing an anchor for these strong-line calibrations at the peak of cosmic star-formation.
\item We find that the MZR at $z \approx\,$1$\,$--$\,$2 decreases to log(O/H) + 12 $\approx$~7.8 $\pm$ 0.1 (15\% solar) at log(M$_{\star}$/M$_{\odot}$)~$\approx$~7.5, without evidence of a turnover or flattening at low stellar masses. The shape of the MZR is consistent with the local relations of \cite{Andrews2013} and \cite{Curti2020} when they are shifted by $\sim$0.3~dex to lower metallicities. This shift is primarily driven by galaxies having lower metallicities at higher redshifts, but also has a secondary dependence on SFR. These results extend the MZR to six times lower in stellar mass and SFR than earlier studies with HST in this redshift range.
\item At $z \approx\,$0.2$\,$--$\,$0.9 our ultra-deep observations reach SFRs that are $\sim$1~dex lower than current calibrations of the MZR from the local Universe. We extrapolate the MZR of \cite{Curti2020} to lower SFRs and find a close agreement with the observations. These findings extend the valid parameter space of the MZR to encompass a range that spans nearly 4~dex in SFR, which will be beneficial for current and future studies utilizing JWST.
\item Our sample at $z \approx\,$0.2$\,$--$\,$0.9 reaches significantly lower SFRs than SDSS galaxies that are well modeled by an FMR that adopts a constant slope ($\gamma$) for the low mass end of the relationship \citep{Curti2020}. Our results align if we relax this constraint and adopt a shallower slope that is consistent with extrapolating to lower SFRs. At $z \approx\,$1$\,$--$\,$2 an MZR with constant $\gamma$ matches our results well, encompassing higher SFR galaxies similar to SDSS. This indicates that future calibrations of the MZR could incorporate $\gamma$ as a function of SFR as an alternative approach to the projection of least scatter.
\item We examined the properties of galaxies in different environments and find at most a tentative $\sim$0.2 dex enhancement in the metallicities of galaxies in groups versus isolated environments. However, the result is not statistically significant because there are a small number of galaxies with individual metallicity measurements between the environments that are also in the same mass range, which is required for a consistent comparison.
\end{enumerate}

In closing, the metallicities of the lowest mass galaxies in our sample approach the low metallicity limits of many current strong-line calibrations. These results together with recent studies utilizing JWST highlight the current need to make direct method measurements at lower stellar masses and SFRs for determining direct metallicities that can be used to extend strong-line calibrations for samples without auroral emission line measurements. These observations reach the lowest masses and SFRs that are accessible to HST, and provide a key benchmark based on established calibrations for comparison with future studies utilizing JWST and Roman.

\acknowledgments

The authors would like to thank the anonymous reviewer for helpful comments that improved the clarity of this paper. M. Revalski thanks Matilde Mingozzi for assistance with UV emission line diagnostics, and Alec S. Hirschauer for useful discussions on spectral stacking and metallicity calibrations. P. Dayal acknowledges support from the NWO grant 016.VIDI.189.162 (``ODIN") and from the European Commission's and University of Groningen's CO-FUND Rosalind Franklin program.

Based on observations with the NASA/ESA Hubble Space Telescope obtained from the MAST Data Archive at the Space Telescope Science Institute, which is operated by the Association of Universities for Research in Astronomy, Incorporated, under NASA contract NAS5-26555. Support for program numbers 15637 and 15968 was provided through a grant from the STScI under NASA contract NAS5-26555. These observations are associated with program numbers \href{https://archive.stsci.edu/proposal_search.php?mission=hst&id=6631}{6631}, \href{https://archive.stsci.edu/proposal_search.php?mission=hst&id=15637}{15637}, and \href{https://archive.stsci.edu/proposal_search.php?mission=hst&id=15968}{15968}. The MUSE portion of this project has received funding from the European Research Council (ERC) under the European Union's Horizon 2020 research and innovation programme (grant agreement No 757535) and by Fondazione Cariplo (grant No 2018-2329). The HST and MUSE data used in this study are available from \cite{https://doi.org/10.17909/q67p-ym16, https://doi.org/10.17909/81fp-2g44, https://doi.org/10.18727/archive/84}. This paper used the photoionization code Cloudy, which can be obtained from \url{http://www.nublado.org} and the Atomic Line List available at \url{http://www.pa.uky.edu/~peter/atomic/}. This research has made use of NASA's Astrophysics Data System.

\facilities{HST (WFC3, WFPC2), VLT (MUSE)}


\software{Astropy \citep{AstropyCollaboration2013, AstropyCollaboration2018, AstropyCollaboration2022}, IPython \citep{Perez2007}, Jupyter \citep{Kluyver2016}, LMFIT \citep{Newville2023}, Matplotlib \citep{Hunter2007, Caswell2021}, NumPy \citep{Harris2020}, Python (\citealp{VanRossum2009}, \url{https://www.python.org}), Scipy \citep{Virtanen2020}}\\

Github repositories cited in this publication:
\begin{enumerate}
\setlength{\itemsep}{-0.25em}
    \item \url{https://github.com/HSTWISP/wisp_analysis/tree/ahenry_mzr}
    \item \url{https://github.com/mrevalski/mudf_analysis}
\end{enumerate}


\bibliography{references}{}
\bibliographystyle{aasjournal}

\appendix
\begin{deluxetable*}{cccccccc}[htb!]
\tabletypesize{\normalsize}
\tablecaption{{\normalsize Quantities Derived from Stacked Spectra}}
\tablehead{
\colhead{log(M/M$_{\odot}$)} & \colhead{N$_\mathrm{gal}$} & \colhead{Mean log(M/M$_{\odot}$)} & \colhead{log(SFR)} & \colhead{E(B-V)} & \colhead{W(H$\beta$)} & \colhead{[O~III] $\lambda$4363/H$\gamma$} & \colhead{12 + log(O/H)}\\[-0.5em]
\colhead{(1)} & \colhead{(2)} & \colhead{(3)} & \colhead{(4)} & \colhead{(5)} & \colhead{(6)} & \colhead{(7)} & \colhead{(8)}
}
\startdata
\textbf{0.23 $\leq$ \textit{z} $\leq$ 0.87} &  &  &  &  &  & \\ \hline
6.5-7.5 & 14 & 7.22 & -1.57 & 0.48 $^{+0.00}_{-0.26}$ & 5.70 $^{+0.00}_{-3.30}$ & 0.03 $^{+0.28}_{-0.00}$ & 8.34 $^{+0.06}_{-0.08}$ \\ \relax 
7.5-8.0 & 12 & 7.76 & -1.10 & 0.00 $^{+0.28}_{-0.00}$ & 5.70 $^{+0.00}_{-3.30}$ & 0.03 $^{+0.32}_{-0.00}$ & 8.46 $^{+0.06}_{-0.06}$ \\ \relax 
8.0-8.5 & 29 & 8.23 & -0.98 & 0.00 $^{+0.12}_{-0.00}$ & 5.70 $^{+0.00}_{-1.80}$ & 0.03 $^{+0.28}_{-0.00}$ & 8.48 $^{+0.03}_{-0.03}$ \\ \relax 
8.5-9.0 & 6 & 8.67 & -0.89 & 0.04 $^{+0.20}_{-0.04}$ & 5.70 $^{+0.00}_{-2.70}$ & 0.03 $^{+0.36}_{-0.00}$ & 8.52 $^{+0.04}_{-0.04}$ \\ \relax 
9.0-9.5 & 5 & 9.28 & -0.37 & 0.04 $^{+0.22}_{-0.04}$ & 5.70 $^{+0.00}_{-2.70}$ & 0.03 $^{+0.36}_{-0.00}$ & 8.54 $^{+0.06}_{-0.03}$ \\ \relax 
9.5-11.0 & 6 & 10.17 & 0.24 & 0.00 $^{+0.30}_{-0.00}$ & 5.10 $^{+0.60}_{-3.00}$ & 0.07 $^{+0.60}_{-0.00}$ & 8.70 $^{+0.08}_{-0.08}$ \\\hline \relax
\textbf{1.00 $\leq$ \textit{z} $\leq$ 2.39} &  &  &  &  &  & \\ \hline
6.5-7.8 & 9 & 7.51 & -0.57 & 0.18 $^{+0.10}_{-0.18}$ & 5.70 $^{+0.00}_{-3.90}$ & 0.14 $^{+0.92}_{-0.00}$ & 7.84 $^{+0.16}_{-0.16}$ \\ \relax 
7.8-8.3 & 19 & 8.11 & -0.15 & 0.00 $^{+0.26}_{-0.00}$ & 0.30 $^{+3.60}_{-0.30}$ & 0.03 $^{+0.36}_{-0.00}$ & 8.10 $^{+0.12}_{-0.16}$ \\ \relax 
8.3-8.8 & 23 & 8.59 & 0.32 & 0.00 $^{+0.26}_{-0.00}$ & 5.70 $^{+0.00}_{-3.60}$ & 0.47 $^{+0.00}_{-0.32}$ & 8.22 $^{+0.08}_{-0.08}$ \\ \relax 
8.8-9.3 & 27 & 9.04 & 0.55 & 0.00 $^{+0.10}_{-0.00}$ & 5.70 $^{+0.00}_{-2.70}$ & 0.22 $^{+0.12}_{-0.16}$ & 8.30 $^{+0.04}_{-0.03}$ \\ \relax 
9.3-9.8 & 22 & 9.56 & 0.88 & 0.14 $^{+0.08}_{-0.12}$ & 5.70 $^{+0.00}_{-3.00}$ & 0.02 $^{+0.24}_{-0.00}$ & 8.34 $^{+0.06}_{-0.04}$ \\ \relax 
9.8-11.0 & 9 & 10.31 & 0.75 & 0.30 $^{+0.16}_{-0.16}$ & 5.70 $^{+0.00}_{-3.30}$ & 0.05 $^{+0.48}_{-0.00}$ & 8.42 $^{+0.06}_{-0.08}$ \relax
\enddata
\tablecomments{The physical quantities derived for stacked spectra over six mass intervals for the low and high redshift samples. Columns list for the galaxies in each bin the: (1) mass range, (2) number of galaxies, (3) mean galaxy stellar mass, and (4) mean SPS-derived SFR in M$_{\odot}$ yr$^{-1}$. The remaining four columns provide the Bayesian inferred (5) extinction due to dust, (6) equivalent width of H$\beta$ in \AA, (7) [O~III] $\lambda$4363/H$\gamma$ ratio, and (8) the gas-phase oxygen abundance. The uncertainties represent one-sigma confidence intervals and uncertainties of zero in one direction indicate that the most likely solution is at the edge of the physically allowed parameter space as described in \cite{Henry2021}.}
\label{tab:results_stacks}
\vspace{-2em}
\end{deluxetable*}
\clearpage
\textcolor{white}{.}
\clearpage

\begin{deluxetable}{rccc}[h]
\tabletypesize{\normalsize}
\tablecaption{Direct Method Metallicities}
\tablehead{\colhead{ID} & \colhead{Redshift} & \colhead{12 + log(O/H)} & \colhead{12 + log(O/H)}\\[-0.5em]
\colhead{} & \colhead{} & \colhead{(Direct Method)} & \colhead{(Strong-line Method)}}
\startdata
1028 & 0.78299 & 7.65 $^{+0.06}_{-0.06}$ & 7.60 $^{+0.45}_{-0.10}$ \\ \relax 
1045 & 0.78695 & 7.54 $^{+0.08}_{-0.08}$ & 7.68 $^{+0.12}_{-0.11}$ \\ \relax 
1495 & 0.78738 & 7.94 $^{+0.14}_{-0.20}$ & 8.32 $^{+0.10}_{-0.10}$ \\ \relax 
1603 & 0.78889 & 8.28 $^{+0.05}_{-0.04}$ & 8.08 $^{+0.10}_{-0.28}$ \\ \relax 
1908 & 0.63634 & 8.39 $^{+0.10}_{-0.12}$ & 8.38 $^{+0.10}_{-0.10}$ \\ \relax 
1950 & 0.68954 & 8.41 $^{+0.14}_{-0.20}$ & 8.20 $^{+0.10}_{-0.11}$ \\ \hline \relax 
856 & 2.32052 & 7.76 $^{+0.20}_{-0.37}$ & 7.90 $^{+0.16}_{-0.16}$ \\ \relax 
1280 & 2.25694 & 8.21 $^{+0.04}_{-0.04}$ & 8.28 $^{+0.10}_{-0.11}$ \\ \relax 
1326 & 2.25484 & 8.04 $^{+0.08}_{-0.09}$ & 8.04 $^{+0.12}_{-0.13}$ \\ \relax 
1449 & 2.26199 & 8.47 $^{+0.10}_{-0.13}$ & 8.36 $^{+0.11}_{-0.11}$ \\ \relax 
1675 & 2.25455 & 8.46 $^{+0.11}_{-0.13}$ & 8.22 $^{+0.12}_{-0.12}$ \\ \relax 
20931 & 2.10019 & 8.10 $^{+0.19}_{-0.25}$ & 8.20 $^{+0.19}_{-0.24}$ \relax 
\enddata
\tablecomments{Metallicities determined using the direct and strong-line methods as shown in Figure~\ref{fig:direct}. The horizontal line separates sources where the direct method metallicity is based on [O~III] $\lambda$4363~\AA\ (upper rows) versus [O~III] $\lambda$1666~\AA\ (lower rows).}
\label{tab:direct}
\end{deluxetable}

\startlongtable
\begin{deluxetable}{ccccc}
\vspace{-3em}
\setlength{\tabcolsep}{0.025in}
\def\arraystretch{0.93}
\tabletypesize{\small}
\tablecaption{Mass-Metallicity Results for Individual Galaxies}
\tablehead{
\colhead{ID} & \colhead{Redshift} & \colhead{Mass (log M$_{\odot}$)} & \colhead{SFR (log M$_{\odot}$ yr$^{-1}$)} & \colhead{12 + log(O/H)}
}
\startdata
501 & 0.44987 & 9.101 $\pm$ 0.100 & -0.441 $\pm$ 0.059 & 8.56 $^{+0.11}_{-0.10}$ \\ \relax 
587 & 0.30800 & 7.179 $\pm$ 0.101 & -0.775 $\pm$ 0.044 & 8.20 $^{+0.10}_{-0.10}$ \\ \relax 
635 & 0.76130 & 9.623 $\pm$ 0.106 & 0.066 $\pm$ 0.030 & 8.70 $^{+0.10}_{-0.10}$ \\ \relax 
656 & 0.76268 & 9.132 $\pm$ 0.101 & -0.368 $\pm$ 0.024 & 8.66 $^{+0.10}_{-0.10}$ \\ \relax 
676 & 0.76138 & 8.200 $\pm$ 0.109 & -0.675 $\pm$ 0.058 & 8.58 $^{+0.10}_{-0.11}$ \\ \relax 
799 & 0.67898 & 8.441 $\pm$ 0.109 & -0.425 $\pm$ 0.048 & 8.46 $^{+0.12}_{-0.10}$ \\ \relax 
827 & 0.30807 & 8.377 $\pm$ 0.101 & -2.223 $\pm$ 0.171 & 8.66 $^{+0.11}_{-0.11}$ \\ \relax 
849 & 0.31984 & 8.002 $\pm$ 0.105 & -2.508 $\pm$ 0.071 & 8.60 $^{+0.10}_{-0.10}$ \\ \relax 
856 & 2.32052 & 8.389 $\pm$ 0.105 & 0.979 $\pm$ 0.039 & 7.90 $^{+0.16}_{-0.16}$ \\ \relax 
897 & 0.67941 & 8.568 $\pm$ 0.101 & -0.178 $\pm$ 0.107 & 8.42 $^{+0.10}_{-0.10}$ \\ \relax 
937 & 0.28816 & 7.497 $\pm$ 0.102 & -2.172 $\pm$ 0.118 & 8.52 $^{+0.11}_{-0.10}$ \\ \relax 
958 & 0.67958 & 8.270 $\pm$ 0.104 & -0.977 $\pm$ 0.076 & 8.52 $^{+0.10}_{-0.11}$ \\ \relax 
964 & 0.67933 & 7.662 $\pm$ 0.109 & -0.424 $\pm$ 0.216 & 8.48 $^{+0.10}_{-0.11}$ \\ \relax 
1023 & 0.78484 & 7.838 $\pm$ 0.101 & -0.715 $\pm$ 0.036 & 7.66 $^{+0.21}_{-0.12}$ \\ \relax 
1028 & 0.78482 & 7.772 $\pm$ 0.100 & -1.098 $\pm$ 0.015 & 7.60 $^{+0.45}_{-0.10}$ \\ \relax 
1038 & 0.86851 & 9.380 $\pm$ 0.112 & 0.039 $\pm$ 0.034 & 8.60 $^{+0.10}_{-0.10}$ \\ \relax 
1045 & 0.78485 & 7.276 $\pm$ 0.101 & -0.485 $\pm$ 0.016 & 7.68 $^{+0.12}_{-0.11}$ \\ \relax 
1069 & 0.78485 & 7.005 $\pm$ 0.143 & -0.493 $\pm$ 0.124 & 7.74 $^{+0.10}_{-0.10}$ \\ \relax 
1074 & 0.67921 & 8.365 $\pm$ 0.103 & -0.681 $\pm$ 0.075 & 8.56 $^{+0.10}_{-0.10}$ \\ \relax 
1082 & 0.68502 & 8.303 $\pm$ 0.111 & -0.695 $\pm$ 0.065 & 8.62 $^{+0.10}_{-0.10}$ \\ \relax 
1121 & 0.25339 & 7.870 $\pm$ 0.101 & -1.874 $\pm$ 0.031 & 8.44 $^{+0.10}_{-0.10}$ \\ \relax 
1195 & 2.25553 & 9.298 $\pm$ 0.118 & 0.671 $\pm$ 0.038 & 8.60 $^{+0.11}_{-0.12}$ \\ \relax 
1240 & 0.68492 & 7.504 $\pm$ 0.105 & -1.105 $\pm$ 0.052 & 8.64 $^{+0.10}_{-0.10}$ \\ \relax 
1264 & 0.68597 & 8.080 $\pm$ 0.104 & -0.539 $\pm$ 0.054 & 8.54 $^{+0.10}_{-0.11}$ \\ \relax 
1275 & 0.68608 & 8.341 $\pm$ 0.115 & -0.954 $\pm$ 0.019 & 8.54 $^{+0.10}_{-0.12}$ \\ \relax 
1280 & 2.25694 & 8.715 $\pm$ 0.102 & 1.367 $\pm$ 0.018 & 8.28 $^{+0.10}_{-0.11}$ \\ \relax 
1287 & 0.79128 & 8.021 $\pm$ 0.104 & -0.342 $\pm$ 0.060 & 8.58 $^{+0.10}_{-0.10}$ \\ \relax 
1292 & 0.68522 & 8.378 $\pm$ 0.105 & -0.987 $\pm$ 0.114 & 8.56 $^{+0.10}_{-0.11}$ \\ \relax 
1302 & 0.67831 & 8.951 $\pm$ 0.109 & 0.184 $\pm$ 0.065 & 8.44 $^{+0.10}_{-0.10}$ \\ \relax 
1326 & 2.25484 & 8.248 $\pm$ 0.107 & 0.849 $\pm$ 0.036 & 8.04 $^{+0.12}_{-0.13}$ \\ \relax 
1354 & 0.67769 & 7.940 $\pm$ 0.102 & -0.839 $\pm$ 0.099 & 8.74 $^{+0.10}_{-0.10}$ \\ \relax 
1364 & 0.67762 & 8.622 $\pm$ 0.102 & -0.802 $\pm$ 0.116 & 8.62 $^{+0.10}_{-0.10}$ \\ \relax 
1401 & 0.30866 & 8.198 $\pm$ 0.100 & -1.441 $\pm$ 0.113 & 8.52 $^{+0.10}_{-0.10}$ \\ \relax 
1403 & 0.68510 & 8.143 $\pm$ 0.104 & -0.489 $\pm$ 0.063 & 8.42 $^{+0.10}_{-0.11}$ \\ \relax 
1414 & 2.21451 & 8.551 $\pm$ 0.166 & -0.076 $\pm$ 0.093 & 8.38 $^{+0.10}_{-0.10}$ \\ \relax 
1440 & 0.30889 & 7.731 $\pm$ 0.103 & -3.512 $\pm$ 0.091 & 8.52 $^{+0.13}_{-0.12}$ \\ \relax 
1449 & 2.26199 & 8.962 $\pm$ 0.110 & 1.256 $\pm$ 0.074 & 8.36 $^{+0.11}_{-0.11}$ \\ \relax 
1462 & 2.04440 & 8.952 $\pm$ 0.120 & 0.549 $\pm$ 0.075 & 8.02 $^{+0.37}_{-0.43}$ \\ \relax 
1467 & 0.65394 & 8.039 $\pm$ 0.123 & -1.325 $\pm$ 0.188 & 8.52 $^{+0.11}_{-0.11}$ \\ \relax 
1470 & 0.42407 & 8.186 $\pm$ 0.102 & -1.788 $\pm$ 0.105 & 8.66 $^{+0.10}_{-0.10}$ \\ \relax 
1495 & 0.78466 & 6.726 $\pm$ 0.110 & -0.659 $\pm$ 0.058 & 8.32 $^{+0.10}_{-0.10}$ \\ \relax 
1574 & 0.31301 & 7.761 $\pm$ 0.104 & -2.268 $\pm$ 0.096 & 8.68 $^{+0.10}_{-0.11}$ \\ \relax 
1576 & 2.31738 & 8.995 $\pm$ 0.107 & 1.273 $\pm$ 0.068 & 8.24 $^{+0.11}_{-0.11}$ \\ \relax 
1603 & 0.78562 & 7.498 $\pm$ 0.102 & 0.148 $\pm$ 0.021 & 8.08 $^{+0.10}_{-0.28}$ \\ \relax 
1609 & 2.22756 & 8.854 $\pm$ 0.105 & 1.472 $\pm$ 0.039 & 8.38 $^{+0.11}_{-0.11}$ \\ \relax 
1611 & 0.40657 & 7.381 $\pm$ 0.102 & -1.924 $\pm$ 0.204 & 8.46 $^{+0.11}_{-0.11}$ \\ \relax 
1627 & 0.28528 & 7.148 $\pm$ 0.106 & -3.519 $\pm$ 0.142 & 8.46 $^{+0.13}_{-0.13}$ \\ \relax 
1675 & 2.25455 & 8.623 $\pm$ 0.112 & 1.138 $\pm$ 0.063 & 8.22 $^{+0.12}_{-0.12}$ \\ \relax 
1704 & 0.65219 & 9.324 $\pm$ 0.103 & -0.613 $\pm$ 0.083 & 8.52 $^{+0.11}_{-0.11}$ \\ \relax 
1730 & 0.64002 & 7.738 $\pm$ 0.110 & -1.407 $\pm$ 0.199 & 8.40 $^{+0.11}_{-0.11}$ \\ \relax 
1773 & 2.29896 & 8.658 $\pm$ 0.128 & 0.853 $\pm$ 0.073 & 8.32 $^{+0.12}_{-0.12}$ \\ \relax 
1797 & 2.25834 & 9.300 $\pm$ 0.172 & 0.290 $\pm$ 0.091 & 8.32 $^{+0.11}_{-0.12}$ \\ \relax 
1800 & 0.76202 & 9.746 $\pm$ 0.101 & 0.413 $\pm$ 0.078 & 8.74 $^{+0.10}_{-0.10}$ \\ \relax 
1803 & 0.79165 & 7.310 $\pm$ 0.133 & -1.454 $\pm$ 0.089 & 8.50 $^{+0.10}_{-0.11}$ \\ \relax 
1831 & 2.34666 & 9.339 $\pm$ 0.122 & 1.127 $\pm$ 0.092 & 8.36 $^{+0.32}_{-0.21}$ \\ \relax 
1908 & 0.63470 & 8.135 $\pm$ 0.102 & -0.396 $\pm$ 0.034 & 8.38 $^{+0.10}_{-0.10}$ \\ \relax 
1934 & 0.65414 & 10.257 $\pm$ 0.102 & 0.725 $\pm$ 0.044 & 8.74 $^{+0.10}_{-0.10}$ \\ \relax 
1950 & 0.67954 & 8.024 $\pm$ 0.101 & 0.681 $\pm$ 0.013 & 8.20 $^{+0.10}_{-0.11}$ \\ \relax 
1988 & 0.65477 & 7.657 $\pm$ 0.103 & -0.325 $\pm$ 0.065 & 8.56 $^{+0.10}_{-0.10}$ \\ \relax 
1997 & 0.76155 & 6.734 $\pm$ 0.131 & -0.621 $\pm$ 0.084 & 7.60 $^{+0.57}_{-0.10}$ \\ \relax 
2094 & 0.28309 & 8.487 $\pm$ 0.100 & -1.315 $\pm$ 0.012 & 8.52 $^{+0.10}_{-0.10}$ \\ \relax 
2169 & 0.30845 & 7.776 $\pm$ 0.101 & -1.556 $\pm$ 0.073 & 8.48 $^{+0.10}_{-0.10}$ \\ \relax 
2226 & 0.83987 & 8.247 $\pm$ 0.118 & -0.369 $\pm$ 0.072 & 8.44 $^{+0.11}_{-0.11}$ \\ \relax 
2438 & 2.37586 & 10.047 $\pm$ 0.175 & 1.025 $\pm$ 0.064 & 8.42 $^{+0.11}_{-0.11}$ \\ \relax 
2691 & 2.26668 & 10.322 $\pm$ 0.118 & 1.766 $\pm$ 0.103 & 8.56 $^{+0.10}_{-0.10}$ \\ \relax 
2742 & 1.99334 & 9.390 $\pm$ 0.170 & 0.554 $\pm$ 0.129 & 8.24 $^{+0.14}_{-0.51}$ \\ \relax 
2754 & 2.25934 & 9.537 $\pm$ 0.157 & 0.503 $\pm$ 0.128 & 8.38 $^{+0.10}_{-0.10}$ \\ \relax 
20253 & 0.25330 & 8.304 $\pm$ 0.100 & -1.473 $\pm$ 0.014 & 8.40 $^{+0.10}_{-0.10}$ \\ \relax 
20570 & 0.67871 & 8.581 $\pm$ 0.107 & -0.982 $\pm$ 0.204 & 8.72 $^{+0.10}_{-0.11}$ \\ \relax 
20931 & 2.10019 & 8.647 $\pm$ 0.119 & 0.870 $\pm$ 0.057 & 8.20 $^{+0.19}_{-0.24}$ \relax
\enddata
\tablecomments{The catalog IDs, spectroscopic redshifts, stellar masses, SFRs, and gas-phase metallicities for individual sources with [O~III] S/N $\geq$ 10 from Figure~\ref{fig:results_individuals}. \textcolor{white}{I'm adding hidden text otherwise latex will crash on this table.}}
\label{tab:results_individuals}
\end{deluxetable}

\begin{figure*}
\centering
\includegraphics[width=0.95\textwidth]{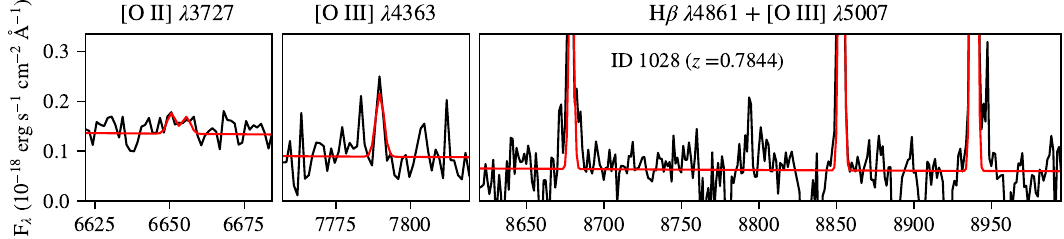}
\includegraphics[width=0.95\textwidth]{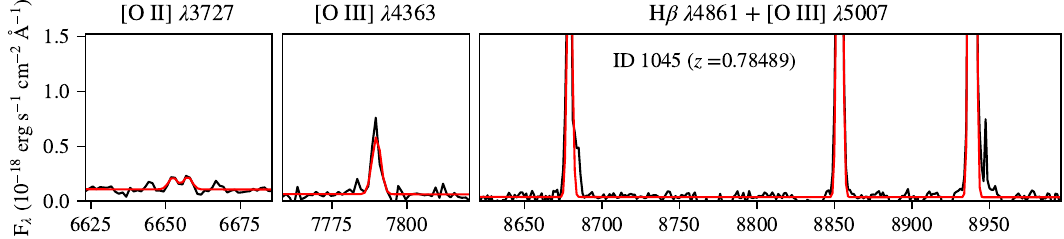}
\includegraphics[width=0.95\textwidth]{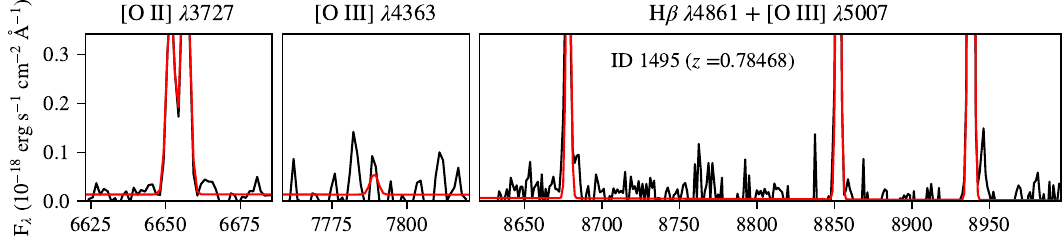}
\includegraphics[width=0.95\textwidth]{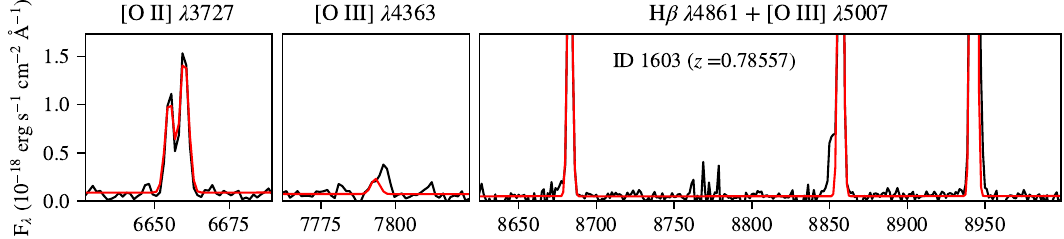}
\includegraphics[width=0.95\textwidth]{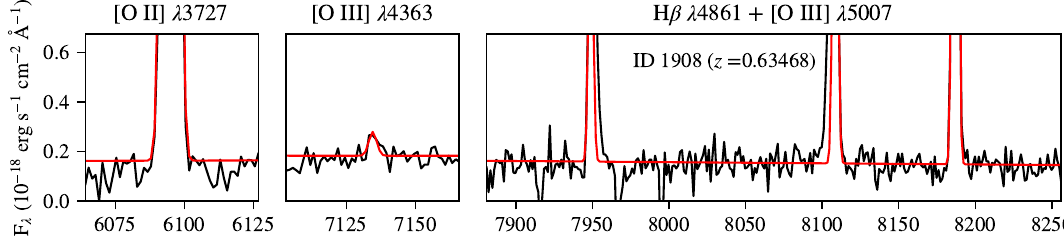}
\includegraphics[width=0.95\textwidth]{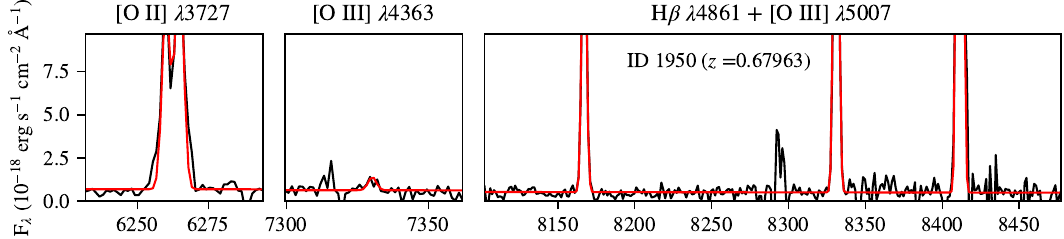}
\caption{Spectral fits for the six objects with [O~III] $\lambda$4363~\AA\ detections used for direct method metallicities (see Figure~\ref{fig:direct}). The panels are shown at observed wavelengths and scaled to half the height of H$\beta$. At high metallicity the nebular [O~II] and [O~III] emission lines are strong, while at low metallicities oxygen is a more important coolant, corresponding to a higher gas temperatures, and so [O~II] is noticeably weaker.}
\label{app:spectra}
\end{figure*}
\addtocounter{figure}{-1}
\begin{figure*}
\centering
\includegraphics[width=0.95\textwidth]{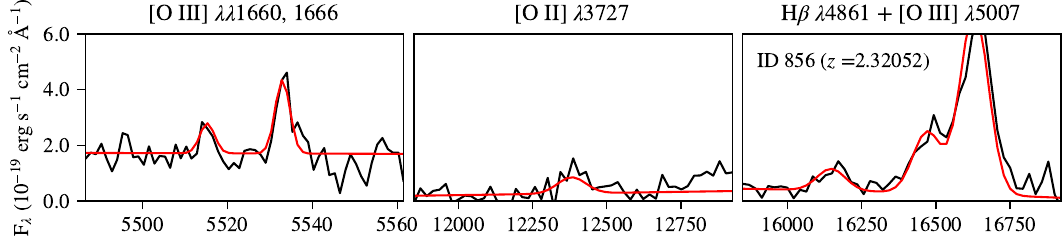}
\includegraphics[width=0.95\textwidth]{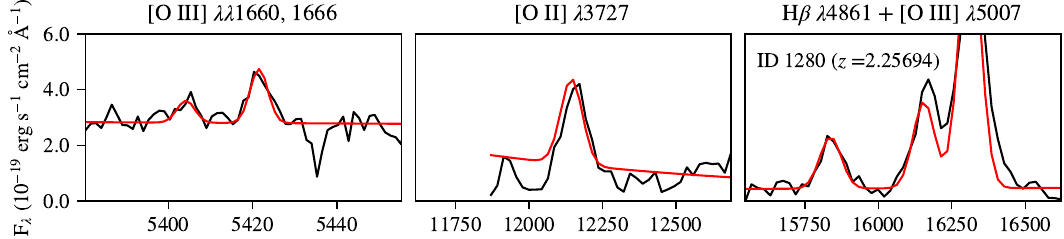}
\includegraphics[width=0.95\textwidth]{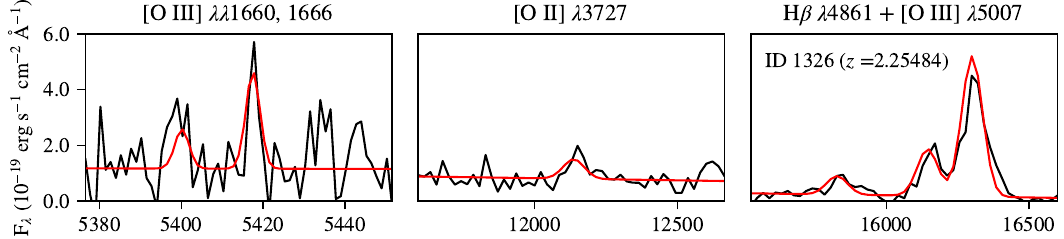}
\includegraphics[width=0.95\textwidth]{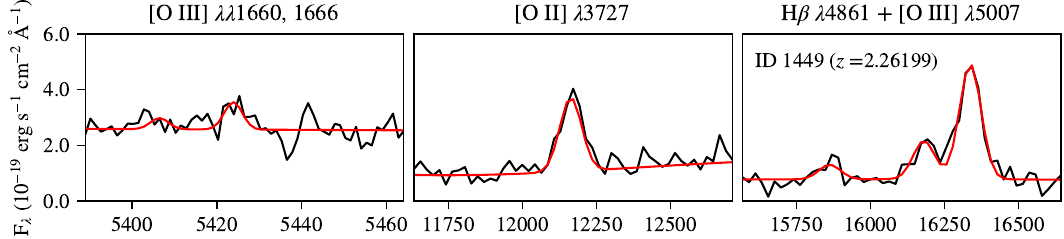}
\includegraphics[width=0.95\textwidth]{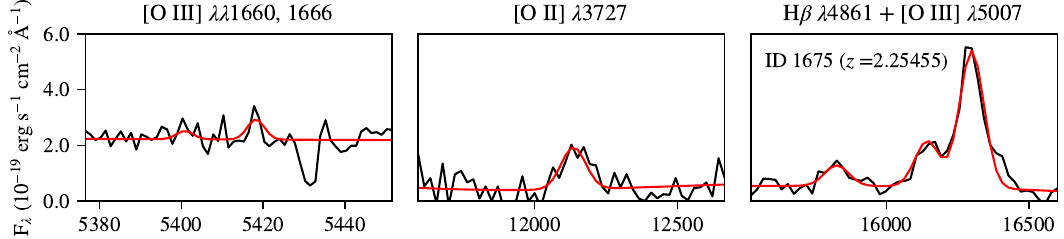}
\includegraphics[width=0.95\textwidth]{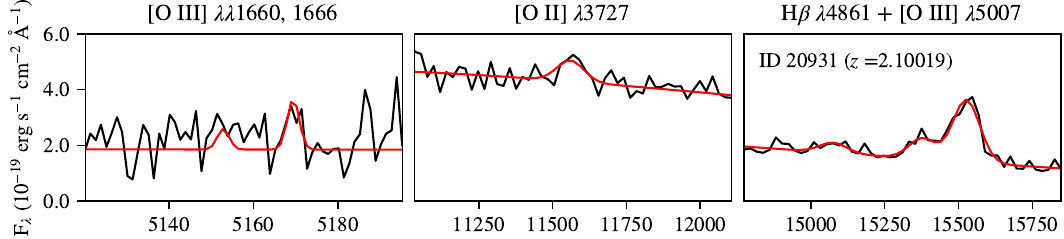}
\caption{\textit{continued.} Spectral fits for the six objects with [O~III] $\lambda\lambda$1660, 1666~\AA\ detections used for direct method metallicities (see Figure~\ref{fig:direct}). The panels are shown at observed wavelengths and are vertically scaled to a fixed flux value of 6.0 $\times$ 10$^{-19}$ erg s$^{-1}$ cm$^{-2}$ \AA$^{-1}$ for comparison.}
\end{figure*}


\end{document}